\documentclass[traditabstract]{aa}
\usepackage{txfonts}
\usepackage{graphicx}
\usepackage{epsfig}
\usepackage{amsmath}
\usepackage{xcolor}
\usepackage{epstopdf}
\usepackage{booktabs}
\usepackage{chngcntr} 
\usepackage{float}

\newcommand{\rb}[1]{\raisebox{1.5ex}[-1.5ex]{#1}}

\begin{document}

\title{Chemodynamical evidence of the HR~1614 moving group\\
as a bar resonance
\thanks{This paper includes data gathered with the 6.5 m Magellan Telescopes located at Las Campanas 
Observatory, Chile.}}

\author{
Andreas J. Koch-Hansen\inst{1} 
\and Vladislav Gulaev\inst{1}
\and Dominik H. Plonka\inst{2}
\and Camilla Juul Hansen\inst{2}
\and Andrew McWilliam\inst{3}
\and Victor P. Debattista\inst{4}
\and Zdenek Prudil\inst{5}
}
\authorrunning{A.J. Koch-Hansen et al.}
\titlerunning{Chemodynamics of the HR~1614 group}
\institute{Zentrum f\"ur Astronomie der Universit\"at Heidelberg, Astronomisches Rechen-Institut, M\"onchhofstr. 12, 69120 Heidelberg, Germany;
\email{andreas.koch@uni-heidelberg.de}
\and Goethe University Frankfurt, Institute for Applied Physics, Max-von-Laue Stra\ss e 12, 60438 Frankfurt am Main, Germany
\and Carnegie Observatories, 813 Santa Barbara St., Pasadena, CA 91101, USA
\and Jeremiah Horrocks Institute, University of Central Lancashire, Preston PR1 2HE, UK
\and European Southern Observatory, Karl-Schwarzschild-Strasse 2, 85748 Garching bei München, Germany
}
\date{}
\abstract{
Moving groups (MGs) are ensembles of disk stars that clump in velocity and action space. A 
multitude 
of explanations as to their origin has been attempted, from dissolved star clusters, to 
{ resonances with the Galactic bar, or other
non-axisymmetric perturbations.} 
The object of this work, 
{ the old HR~1614 group, 
has in the past already been shown to not be a disrupted cluster in terms of its broad age and metallicity distribution. 
Thus its stars have more likely been trapped by the Milky Way bar at corotation}. 
{ Here, we  present a new study that aims to chemically and dynamically characterize this group in the context of such a resonance}.
{ To this end, we used} MIKE/Magellan high-resolution, high signal-to-noise spectra of six member stars of the
 HR~1614 MG { to determine} the
chemical abundance ratios for
24 elements, several for the first time in those stars. 
The Fe-abundances of our sample range from +0.13 to 0.38 dex 
and the majority of abundance ratios is fully in line with those of metal-rich Milky Way disk stars, with few
exceptions.
In particular, the group's eponym, HR~1614, is enhanced in essentially all abundances and coincides with the higher-[X/Fe] trends of the Milky Way's
thick disk.
All elements (but S and Ca) show significant intrinsic abundance scatter that argue against this 
MG having formed in a contained environment such as a dissolved cluster.
Isochrone ages, tailored to the measured metallicities 
support earlier findings of a broad age mix in this group, ranging from $\sim$1 Gyr 
to as old as $\sim$8 Gyr. 
We performed 
an orbital frequency analysis in a 
Galactic potential that includes a rotating bar. 
Indeed, our sample stars fall onto thin resonances on 
corotation.
Hence, our results support the idea that the HR~1614 MG is a resonantly perturbed 
disk feature, mustering a m\'elange of stars from different parent populations in the inner Galaxy.
}
\keywords{Stars: abundances --- Stars: kinematics and dynamics --- Galaxy: disks --- Galaxy: kinematics and dynamics --- open clusters and associations: individual: HR 1614 }
\maketitle
%
%
%
\nolinenumbers
\section{Introduction}
{\em Via lactea est omnis divisa in partes tres.} -- this dictum, loosely adapted from G.I. Caesar (58 B.C.), 
obviously purveys a hopelessly antiquated 
concept of a three-component (disk/bulge/halo) Milky Way (MW) and its siblings \citep[e.g.,][]{Kordopatis2013}. 
Evidence has been emerging that each of these is subdivided into thin/thick disks
of various metallicities \citep[e.g.,][]{Gilmore1983,Kordopatis2013,Bensby2014}, inner/outer halos displaying abundance gradients  
\citep[e.g.,][]{Carollo2007,Koch2008M31}, and a complicated bulge plus bar structure \citep[e.g.,][]{Blitz1991,Stanek1997,McWilliam2010,Wegg2013,Kunder2020}. 
Separating all these is not straightforward and further complicated by dynamical effects, dislocating stars and clusters from their birth radii and redistributing angular momentum. 
This can lead to overlapping stellar populations \citep[e.g.,][]{Koch2018Gaia,Gran2021} 
and various mixing effects, such as  radial migration, blurring, or churning, all of which are essentially resonant effects due to 
non-axisymmetric perturbations \citep[e.g.,][]{Sellwood2002,Haywood2008,Frankel2020}.
One class of objects attesting to the dynamic (co-)evolution of these substructures are moving groups (MGs), 
i.e., ensembles of stars that are distinct in velocity and action space \citep{Eggen1958,Eggen1965,Trick2021}.

The HR\,1614 MG was initially characterized as an old ($\sim$2 Gyr), comoving  system in the survey of some 400  stars by \citet{Eggen1971}, who later noted the supersolar 
metallicities of 
the closest member (HR~1614$\equiv$Hip~23311) and 25 bright associates \citep{Eggen1978}. 
Later, the group's age was revised upwards to as high as $\sim$5 Gyr  \citep{Eggen1992}. 
\citet{Feltzing2000} re-examined the HR~1614 group and its origin using various available data sets and broadly sampling  phase space.
In total,  44 stars were identified as possible members of the MG, with 60\% being more metal rich than the Sun, at a mean of $\sim$0.2 dex  
and an age back to Eggen's original assessment of $\sim$2 Gyr. 
\citet{Kushniruk2020} revisited the group from a legion of surveys resulting in a much larger sample of member candidates.
Their ensuing isochrone fits
revealed the presence of multiple stellar populations within the MG with ages between 2 and 8 Gyr and with a large spread in metallicity. 
In consequence, the origin of the HR~1614 MG as a dissolved (open) star cluster was discarded \citep[cf.][]{Gagne2021}. 

Dynamically, its members stand out in velocity (U,V)-space at $\sim$(0, 60) km\,s$^{-1}$ and within V$\pm$10 km\,s$^{-1}$
of the reference star HR~1614. 
Given their propinquity and the tilt in (U,V)-space (see also \citealt{Ramos2018}), { the HR~1614 MG has been identified as being part of the Hercules stream, 
which, in turn, constitutes a corotation (CR) or outer Lindblad resonance (OLR) with the Galactic bar} \citep{Dehnen2000,Hunt2018,DOnghia2020,Chiba2021,Li2024a,Li2024b}.

As to a detailed chemical inventory, \citet{DeSilva2007} conducted the first  high-resolution spectroscopic 
study on 18 of the members of the group. Their results confirmed that the HR 1614 MG is a 2 Gyr old, metal-rich system with [Fe/H] $\sim$0.25 dex. 
Additionally, chemical homogeneity due to the small scatter of only about 0.01 dex for several elements coupled with the purported 
contemporaneous formation of the member stars led to their  conclusion that this MG originated from a dispersed star forming event \citep[see also][]{Antoja2008}. 
However, the main objective of the work of \citet{DeSilva2007} was to perform a differential abundance study relative to the reference star HR~1614
 in order to gauge the abundance scatter with respect to  the measurement errors. Accordingly, only a brief discussion of the ages and kinematics of the stars was given, and
 the group's overall abundance patterns were also not placed in the broader picture of the underlying MW components.
 { Similarly, while \citet{Kushniruk2020} identified a large number of member candidates with this MG, no detailed chemo-dynamic interpretation was carried out for those stars. 
In the present work, we aim to fill in this gap and to provide a detailed chemical abundance, age, and orbital analysis of a number of the HR~1614 MG's member stars
in order to connect it with the underlying MW components and to 
clarify its dynamic origin by resonant bar trapping.
}

This paper is organized as follows: in Sect.~2 we introduce the spectroscopic data. The abundance analysis is described in Sect.~3 and its results presented in Sect.~4. 
We derive stellar ages in Sect.~5, and
in Sect.~6, the investigation of the group's dynamics is performed before concluding with a discussion in Sect.~7. 
\section{Observations and data reduction}
Six stars that are believed to be part of the HR 1614 MG were selected from the sample of \citet{DeSilva2007}. Our choice of those stars was
simply based on their brightness and observability. 
Our sample was observed with the Magellan Inamori Kyocera Echelle (MIKE) spectrograph at the 6.5 m Magellan 2  Clay telescope at the Las Campanas Observatory in Chile during the night of October 16, 2016. 
 Each star was observed over 
four to five individual exposures with exposure times of 30 to 300~s each to facilitate cosmic ray removal. The seeing was overall unstable, varying   
between 2.0$\arcsec$ and 3.2$\arcsec$. We chose the slit width of 0.5$\arcsec$, yielding a  resolving power  of around  40000. The full wavelength range of the red arm of the spectrograph that we employ here 
is 4900--9500 \AA. 
The signal-to-noise ratio for the bright targets ranges from 150 per pixel at shorter wavelengths to 300 per pixel in the red. 
The data reduction process followed standard procedures \citep{Kelson2000,Kelson2003}, including bias correction, flat fielding, scattered light removal, spectral extraction, 
radial velocity correction (see Sect.~6), and 
continuum normalization 
(see \citealt{KochHansen1261} for a data set procured in the same night and processed in the identical manner).
\section{Abundance analysis}
We performed a standard abundance analysis that is based on both equivalent width (EW) measurements and spectral synthesis, where appropriate. 
For the major part, we restricted the EW determinations to  the wavelength range between 5000 and  6600 \AA; redwards of  6600 \AA , telluric absorption 
began hampering the measurements, whereas lines below $\sim$5000 \AA~become progressively 
blended, complicating the proper continuum placement in our metal-rich stars. 
We utilized the line list of \citet{Ruchti2013}, which was constructed for stars of  solar metallicity, and  fitted Gaussian line profiles to the unsaturated lines with IRAF's {\em splot} task.%
\begin{table*}[htb!]
\caption{
Linelist
}
\centering
\begin{tabular}{cccccccccc}
\hline\hline       
 & $\lambda_0$  & EP & & \multicolumn{6}{c}{EW [m\AA]} \\
\cline{5-10}
\rb{Species} &  [\AA] & [eV] &\rb{log\,$gf$}  & Hip\,11575 & Hip\,22336 & Hip\,22940 & Hip\,23311 & Hip\,26834 & Hip\,9353 \\
\hline       
26.0 & 5001.86 &  3.88 &  $-$0.010 &    \phantom{1}97 &   100 &   130 &   106 &   134 &    \phantom{1}90 \\
26.0 & 5044.21 &  2.85 &  $-$2.038 &   \phantom{1}81 &    \phantom{1}76 &    \phantom{1}93 &   144 &    \phantom{1}88 &    \phantom{1}75 \\
\hline                  
\end{tabular}
\tablefoot{Table~1 is available in its entirety in electronic form via CDS. A portion is shown here for guidance regarding content and form.}
\end{table*}
We accounted for hyperfine splitting (chiefly for the Fe-peak elements), except for the weakest lines, using data from 
 \citet{Johnson2006,DenHartog2011,Mashonkina2014,Lawler2015,Lawler2019,Shi2018}.
 A portion of our line list is given in Table~1. 

Our study has 55 lines in common\footnote{Unfortunately, \citet{DeSilva2007} do not list any EWs for their other stars, but only one object -- HR~1614, which constitutes the reference star for their differential analysis.} with the analysis of \citet{DeSilva2007} of Hip~23311.
From these, we find an excellent agreement in our EW measurements with a mean deviation of a mere (0.8$\pm$1.3) m\AA~with a 1$\sigma$-scatter of 9.8 m\AA. The corresponding fractional error is as low as (2$\pm$2)\%
and the two largest outliers, by $\sim$30 m\AA~(or $\sim$30--40\%),  are a Ni and a Mg line. 

The actual abundance derivation was carried 
out 
following the overall methods used in, e.g., 
\citet{Koch5897}, 
with the MOOG abundance code (\citealt{Sneden1973}; 
2019 version), building on the 72-layer, one-dimensional, plane-parallel grid of 
ATLAS atmospheres with Solar-scaled opactiy distributions, {\sc ODFNEW} \citep{CastelliKurucz2003}, 
under the assumption of Local Thermodynamic Equilibrium (LTE) for all species. 
\subsection{Stellar parameters}
Initial values for the stellar effective temperature, T$_{\rm eff}$, were obtained from the $B_2$ and $V_1$ Geneva photometry of \citet{Paunzen2022}, 
where we used the reddening maps and extinction law of 
\citet{Schlafly2011,Green2019} and applied the color-T$_{\rm eff}$-calibration of  \citet{RamirezMelendez2005}. 
One exception is Hip 23311, for which we resorted to the early photometry of \citet{Eggen1977}.
For the surface gravity, log\,$g$, we commenced with the 
canonical  stellar relation from the stars' temperature, dereddened magnitude, distance (in turn based on Gaia parallax, \citealt{BailerJones2018}), and mass.  
The microturbulent velocity, $v_{\rm mic}$, was initially set to 1.5 km\,s$^{-1}$, an initial mass of 0.8 M$_{\odot}$ was adopted (see also \citealt{DeSilva2007}), and we started from Solar metallicities. 

As the next step, T$_{\rm eff}$ was refined by using about 100 iron lines and balancing out the abundance with respect to the excitation potential. 
Simultaneously, $v_{\rm mic}$ was adapted until the iron abundance showed no trend with the reduced width RW$=\log\left({\rm EW}/\lambda\right)$. 
The stellar surface gravities were then derived by imposing ionization equilibrium
between Fe\,{\sc i} and Fe\,{\sc ii} lines.
All parameters were thus iterated towards convergence, where we used the last derived iron abundance from the neutral lines as input to the next iteration's metallicity. 
In all this, too strong and very weak lines with  RWs above  $-$4.5 and below $-$5.5 were culled from the analysis.
The final stellar parameters used for the remainder of this study are listed in Table~2.
\begin{table*}[htb!]
\caption{
Stellar parameters. %
}
\centering
\begin{tabular}{ccccccccc}
\hline\hline       
 & ID &  &  & T$_{\rm eff}$(spec) & T$_{\rm eff}$($B_2-V_1$)  & & v$_{\rm mic}$  & [Fe/H] \\
\rb{Nr.} & (Hip) & \rb{V$_0$} & \rb{$B_2-V_1$} & [K] &  [K] & \rb{log\,$g$} & [km\,s$^{-1}$] & dex \\ 
\hline       
1 & 11575 & 7.93 & 0.395 & 5900 & 5733  & 3.70 & 1.15 & 0.20 \\ 
2 & \phantom{2}9353 & 5.81 &  0.349 & 6100 & 6279  &  3.95 & 1.10 & 0.32  \\ 
3 & 22336 & 5.37 &  0.373 & 5900 & 6468  & 3.90 & 0.80 & 0.30  \\ 
4 & 22940 & 8.29 &  0.524 & 5300 & 5188  & 4.33 & 0.40  & 0.17 \\
5 & 23311 & 6.01 &  \ldots & 4850 & \ldots & 4.23 & 0.90 & 0.13  \\ 
6 & 26834 & 7.15 &  0.477 & 5575 & 5876  & 4.20 & 0.60  & 0.38 \\ 
\hline                  
\end{tabular}
\end{table*}
\subsection{Differential analysis}
A differential analysis relative to a reference star of similar parameters often proves useful to alleviate
systematics in abundance determinations, mainly 
in terms of uncertain $\log\,gf$ values and the 
adopted parameter scales 
\citep{Fulbright2006,Koch2008}. 
We investigated this option for our study with a line-by-line 
differential analysis relative to the Sun,
exemplified 
for star Hip~22336, with its parameters closest to the Sun.
To this end, we employed the Solar line list from \citet{Fulbright2006}
and measured the same lines in the target star. Solar parameters were adopted from \citet{Fulbright2006}, 
namely (T$_{\rm eff}$, log\,$g$, v$_{\rm mic})_{\odot}$ = 
(5770 K, 4.44 dex, 0.93 km\,s$^{-1}$). 
In order to recover excitation and RW balances of the line-by-line 
differences $\epsilon_{\ast}-\epsilon_{\odot}$, 
the ``absolute'' parameters for this star (see above and Table~2)
had to be altered, albeit to within their uncertainties. The resulting
differential parameters for Hip~22336 are thus
(T$_{\rm eff}$, log\,$g$, v$_{\rm mic})_{\ast}$ = (5930 K, 4.12 dex, 1.10 
km\,s$^{-1})$. Likewise, we find a differential Fe-abundance slightly lower, at 0.23$\pm$0.01 dex, for both neutral and ionized species. 
Given the broad mix of parameters across the sample and since the 
results are in broad agreement, we will not pursue this method 
any further.
\subsection{Error analysis}
An order of magnitude for the statistical 1$\sigma$-error on  T$_{\rm eff}$ and v$_{\rm mic}$ was estimated by 
adjusting these parameters until the slopes of the Fe-abundance  were still within the 1$\sigma$-scatter in the excitation potential and RW-diagrams, respectively.
This yields formal uncertainties of $\pm$30 K and $\pm$0.1 km\,s$^{-1}$. 
Moreover, we note that the mean deviation between the  spectroscopic and photometric temperatures 
is 33$\pm$115 K, which alters the abundance results only marginally.
While agreeing very well on average, the 1$\sigma$-scatter in the comparison with the T$_{\rm eff}$-scale of \citet{DeSilva2007} is larger, at 160 K. In the following, we
will adopt a representative temperature uncertainty of 100 K on our measurements. 
Likewise, the star-by-star comparison with the literature values reveals a fair agreement between the gravities with the literature values being higher by 0.25 dex at a 1$\sigma$-scatter of 0.14 dex. 
Finally, the microturbulent velocities are broadly in agreement with those of \citet{DeSilva2007}, with a spread of 0.19 km\,s$^{-1}$. 

The main sources for statistical abundance errors are from the measurements of EWs per continuum placement and the number of used lines, $N$.
Therefore, we state on our following results the statistical errors via the standard deviation of the abundance values and the number of used lines. 

Systematic errors, in turn, were derived by varying every stellar parameter about its uncertainty (estimated above) individually, computing new abundances and noting the difference to the unperturbed 
values. Furthermore, an additional analysis with $\alpha$-enhanced model atmospheres ({\sc AODFNEW}) was performed, thereby mimicking a halo-like [$\alpha$/Fe] ratio of 0.4 dex, which reflects our {\em ab initio} ignorance of the stars' $\alpha$-abundances. 
The final systematic error is calculated from the squared sum of those differences.
However, this should be considered as an upper limit, since cross-terms between stellar parameters were not taken into account \citep{McWilliam1995}. 
The results are given in Table ~A.1 in the appendix, 
for the example of target star Hip~11575. 
\section{Abundance results}
The results of the chemical abundance analysis for the entire sample are presented in 
Table~A.2 in the appendix, where abundances from ionized species are given relative to ionized iron, and neutral species relative to Fe\,{\sc i}. 
Throughout, we adopted the Solar abundances from \citet{Asplund2009}.
In the following, our measurements are placed in context with the Galactic  thin and thick disks,  with the literature data indicated in the respective captions
(Figs.~1--7). We also note that all literature abundances are reported 
in LTE unless noted otherwise. 
\subsection{Iron}
All in all, the iron abundance is supersolar for all stars with a mean of 0.25 dex and a 1$\sigma$-scatter of 0.09 dex. This lies well above the typical spread in a given cluster or formerly coherent 
and now defunct structure, which tend to have intrinsic spreads well below 0.02 dex for most elements  \citep{Sinha2024}, and again argues against  the HR1614 MG being a dissolved star cluster
\citep{Kushniruk2020,Li2024a}. 
For our stars, the values are in excellent agreement to within 0.10 dex with those of  \citet{DeSilva2007}, 
differing only by $\sim$0.14 dex for the eponymous HR 1614 ($\equiv$Hip 23311). 
The latter star had been analysed by \citet{Antipova2016}, who found it to be rather ``anomalous'' in the sense of its stellar parameters strongly differing between photometry and spectroscopy, also noting several  apparent abundance anomalies.
This finding led those authors to conclude that Hip 23311 is a red dwarf with strong, solar-like atmospheric activity.
We also note that, per construction, ionization balance is fulfilled in the stars, with a mean [Fe\,{\sc i}/{\sc ii}] ratio of 0.00$\pm$0.01 dex. 
\subsection{Light elements: C, O, Na, Al, K}
We determined the carbon abundance in all six stars from the two atomic lines at 5380 and 6587 \AA.  Corrections for Non-LTE (NLTE) 
were interpolated from \citet{Amarsi2019} and are typically less than 0.1 dex.
Our stars lie, within the uncertainties, on the roughly flat trend delineated by similar MW field stars at Solar to supersolar metallicities. 
Here, and in many other elements discussed below, Hip~23311 is an outlier with a carbon-enhancement of $\sim$0.3 dex.

Oxygen abundances are based on the triplet lines around 7774 \AA. Contrary to carbon, the [O/Fe] ratios are prone to large NLTE corrections. 
In Fig.~1, we show our results together with the uncorrected (left panels) disk star sample of \citet[][and references therein]{Amarsi2019} and their sample 
after applying their 3D-NLTE corrections (right panel; see also \citealt{Sitnova2013,Steffen2015}). In either case, the HR~1614 MG stars are fully consistent with the thin disk trend,
even tighter in the corrected case, with the exception of Hip~23311, which shows an enhancement of [O/Fe]$\sim$0.2 dex. This value is in agreement with the 
higher-$\alpha$ sequence that is populated by thick disk stars \citep[e.g.,][]{Franchini2021}. We note that, despite this star's still relatively high metallicity of $\sim$0.1 dex, it is the most ``metal-poor'' object in our sample.

\begin{figure}[htb!]
\begin{center}
\includegraphics[angle=0,width=1\hsize]{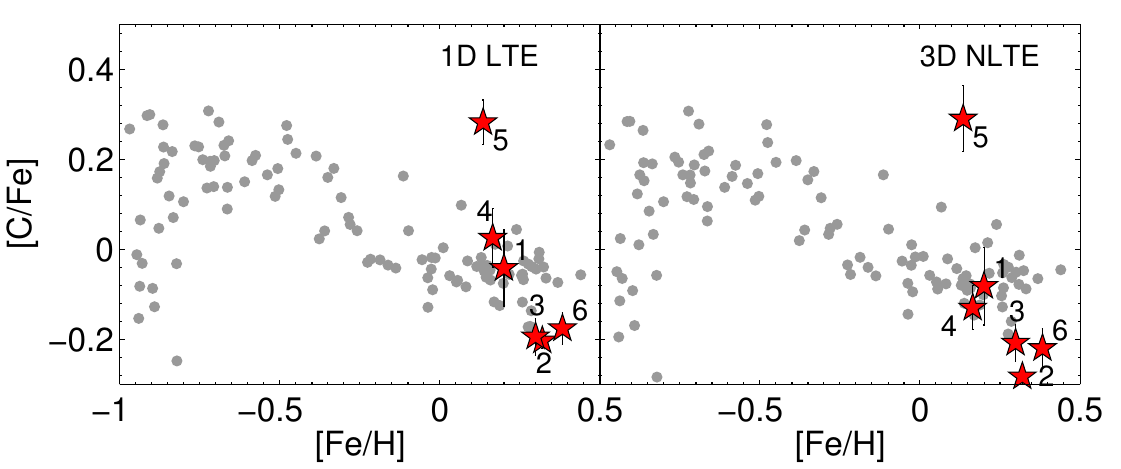}
\includegraphics[angle=0,width=1\hsize]{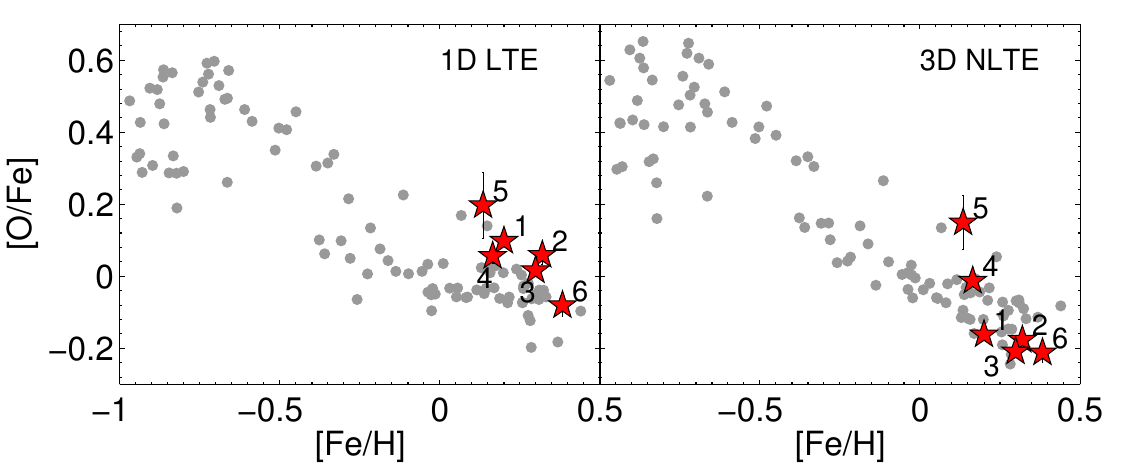}
\end{center}
\caption{Carbon (top panels) and oxygen (bottom) abundances in 1D LTE and after applying NLTE 3D corrections. The corrections and literature data are from \citet{Amarsi2019}. The most metal-`poor' star of our sample,
Hip 23311, has an elevated ratio, in line with a thick disk origin. In this and the following figures, numbers encode
the target stars as follows: 1: Hip~11575; 2: Hip~9353; 3: Hip~22336; 4: Hip~22940; 
5: Hip~23311; 6: Hip~26834.}
\end{figure}

Fig.~2 shows the abundance ratios of the remaining light elements with literature data for MW disk stars as indicated in the caption. 
Na abundances were derived from the line pairs at $\sim$5682 and $\sim$6154 \AA. 
\begin{figure}[htb!]
\begin{center}
\includegraphics[angle=0,width=1\hsize]{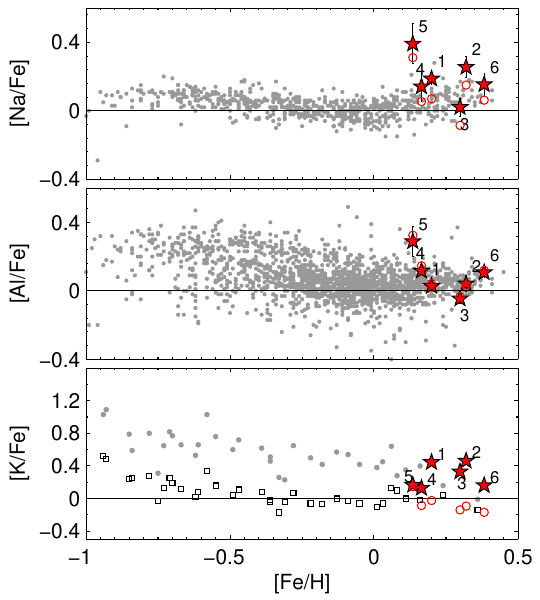}
\end{center}
\caption{Abundance ratios of the light elements Na, Al, and K for our target stars (red star symbols) and literature data (gray dots), which have been taken from 
\citet{Takeda2002},  
\citet{Bensby2014}, 
and \cite{DelgadoMena2017}. 
We also show NLTE-corrected values (open circles) with corrections from \citet{Lind2011}, \citet{Nordlander2017}, and \citet{Takeda2002}. The bottom panel also includes NLTE abundances for the disk stars (black open squares) as 
provided by \citet{Takeda2002}.}
\end{figure}
Most stars follow the disk trends for metal-rich stars, even more so after adjusting the abundances by  $\sim$0.1 dex for each star to account for 
the NLTE corrections implied by \citet{Lind2011}.
We determined the aluminium abundances of our stars from the moderately strong 6696, 6698, and 7835 \AA~lines, and 
determined small NLTE corrections \citep{Nordlander2017}. 
For potassium, 
only one line at 7698 \AA~was useable, which  was not hampered by the overlapping telluric A-band. Also for K, our sample is fully in line with 
disk stars from the literature \citep{Takeda2002}. The latter work's NLTE corrections lower the [K/Fe] ratios to the Solar level over a broad range in metallicity; similar results are achieved 
when using the corrections of \citet{Reggiani2019}. 
In all these elements but K, Hip~23311 stands out in that it has enhanced abundance ratios.
\subsection{$\alpha$-elements: Mg, Si, S, Ca, Ti}
The abundance results for the $\alpha$-elements are chiefly based on standard line lists and procedures as in our previous works. 
The element sulfur, less studied, was analyzed by synthesizing the M6 and M8 multiplets, using  line data from \citet{Caffau2005}. 
Fig.~3 shows our abundance ratios on the trends for MW thin and thick disk stars, with reference data from \citet{Caffau2005} and \citet{Bensby2014}.
\begin{figure}[htb!]
\begin{center}
\includegraphics[angle=0,width=1\hsize]{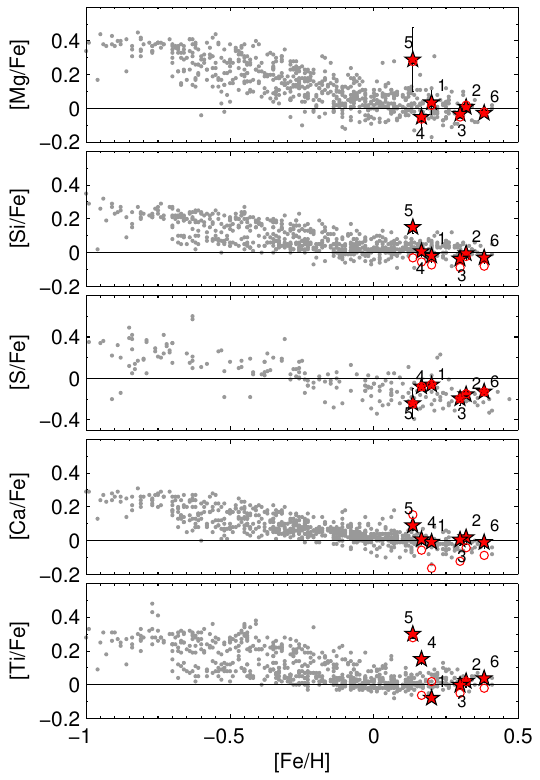}
\end{center}
\caption{[X/Fe] for various $\alpha$-elements of the HR 1614 MG stars (red symbols) compared to the MW disks (gray dots), using data from \citet{Bensby2014} and \citet{Caffau2005} for sulfur.
All [X/Fe] and [Fe/H] are LTE values, while red open circles 
include NLTE corrections in the elements' abundances for our six stars.}
\end{figure}
For completeness, we investigate the effects of NLTE on all elements in this group using line-by-line data for   
Mg \citep{Bergemann2017},  
Si \citep{Bergemann2013}, 
Ca \citep{Mashonkina2007}, and  
Ti \citep{Bergemann2011}. 
The corrections for sulfur are small in Solar type stars
\citep{Amarsi2025}, but also strongly hinge on the treatment of 
the cross section of hydrogen collisions
\citep{Koch2011Sulphur}. All NLTE-corrected values are indicated as open red circles in Fig.~3 and the following figures.

One point to note are the error bars for Mg (top panel of Fig.~3). 
Magnesium lines are typically very strong, especially the ones near 5528 \AA~and 5711 \AA. 
In order to assess the risk of possible saturation, we synthesized those two lines, noting the effect that varying abundances 
had on the line fit. The poorest result was thus obtained for the most metal-rich star in our sample, Hip~26834. 
Overall, the [$\alpha$/Fe] ratios agree well with the underlying disk trends, again with the exception of Hip~23311. Similarly, 
the largest scatter over the metallicity range covered by our small sample is seen for Ti (bottom panel of Fig.~3). 
We note, however, that this is again driven by the two most metal-poor stars in our sample that show [Ti/Fe] ratios enhanced to 0.3 and 0.2 dex above the, otherwise, 
Solar ratios in our stars. 

In order to emphasize the contrast in our data, we show in Fig.~4 the straight average over the four main $\alpha$-elements Mg, Si, Ca, and Ti, disregarding 
differences in their nucleosynthetic formation details due to their hydrostatic or explosive nature 
\citep{Woosley1995,Venn2004}. 
\begin{figure}[htb!]
\begin{center}
\includegraphics[angle=0,width=1\hsize]{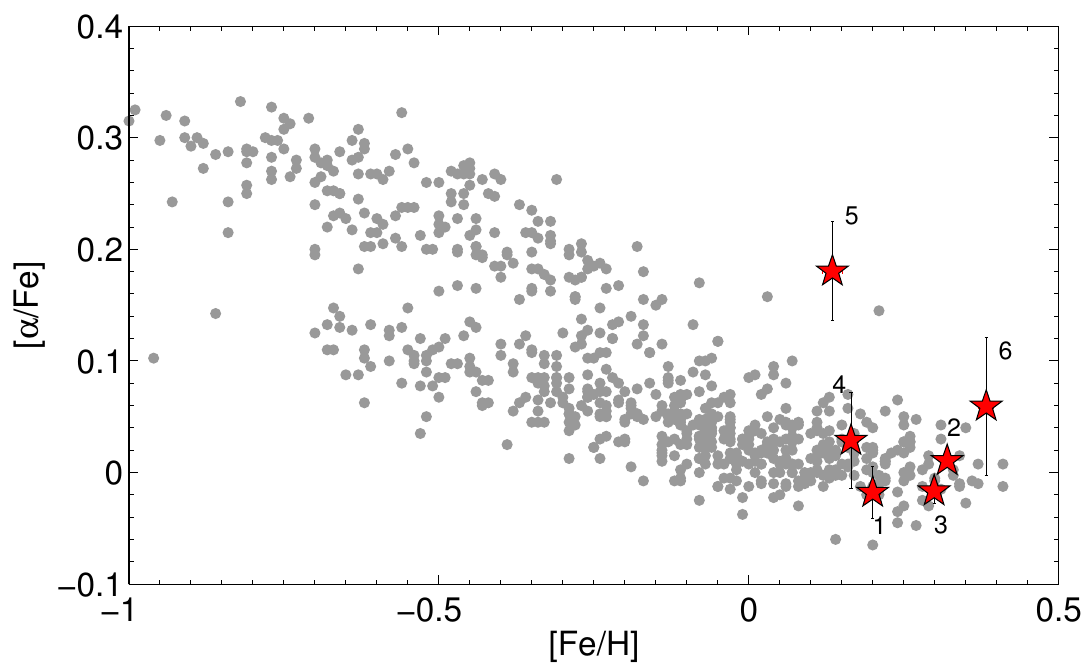}
\end{center}
\caption{Mean [$\alpha$/Fe] ratio of the elements Mg, Si, Ca, and Ti in LTE; 
oxygen and sulfur were not included in this average.
Literature data are from \citet{Bensby2014}. Symbols are as in Fig.~3.}
\end{figure}
All but one star match the literature values for metal-rich thin disk stars. 
As before, it is Hip~23311 (and, to a lesser extent, Hip~26384) that deviates from the Solar (averaged) abundance ratios and rather follows the rising trend
of the higher-$\alpha$ stars in the thick disk.
Thus our chemical evidence so far shows that 
there is a mixed origin of the HR~1614 MG of both thin and thick disk stars. 
Indeed, large local samples show 
chemically thick disk stars 
among the 
kinematically thin disk population, 
and vice versa 
(e.g., \citealt{Bensby2014,McWilliam2016}; and Sect.~6.2). 
 In this regard, Hip~23311 and Hip~26384 are not unusual. However, the iron abundance for  
 Hip~23311 in our work is lower by $\sim$0.2 dex compared to the value reported by 
 \citet{DeSilva2007}.
\subsection{Fe-peak elements: Sc, V, Cr, Mn, Co, Ni, Cu}
In Figs.~5 and 6 we show our results for the Fe-peak results in comparison with literature data as indicated.
\begin{figure}[htb!]
\begin{center}
\includegraphics[angle=0,width=1\hsize]{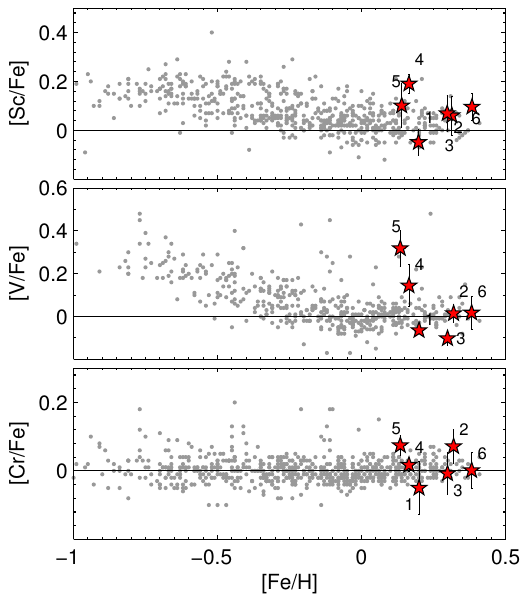}
\end{center}
\caption{Same as Fig.~3, but for the Fe-peak elements Sc, V, and Cr. Literature data are from \citet{Bensby2014} and \citet{Battistini2015}.}
\end{figure}
None of Sc, V, and Ni were corrected for departures from LTE
\citep{Nissen2024}. 
The elements Sc, V, and Cr exhibit trends that are in line with the underlying MW distribution. 
Here, the [Sc/Fe]-ratios are very close to typical MW stars for their [Fe/H].  
While Type Ia Supernovae (SNe Ia; including those at sub-Chandrasekhar mass) tend to  under-produce Sc 
in the thick disk,  the
[Sc/Fe] ratios are enhanced, often in lockstep with the elevated
$\alpha$-levels found in systems with 
enhanced SNe II ejecta \citep[e.g.,][]{Prochaska2000}. 
Thus, the solar-like [Sc/Fe] ratios in the HR~1614 MG 
stars suggest a similar SNe II/Ia ratio as in the Solar neighborhood.
The same trend is visible for [V/Fe], albeit noisier, which is mainly driven by Hip~23311, but a similar scatter is seen in  the values reported for disk stars in the literature. 
 
For Cr, we applied NLTE corrections for those lines with available data  \citep{Bergemann2010Cr}. 
However, those corrections largely average out amongst the used lines for our targets, 
and the only discernible result is a slightly increased line-to-line scatter for each star.

\begin{figure}[htb!]
\begin{center}
\includegraphics[angle=0,width=1\hsize]{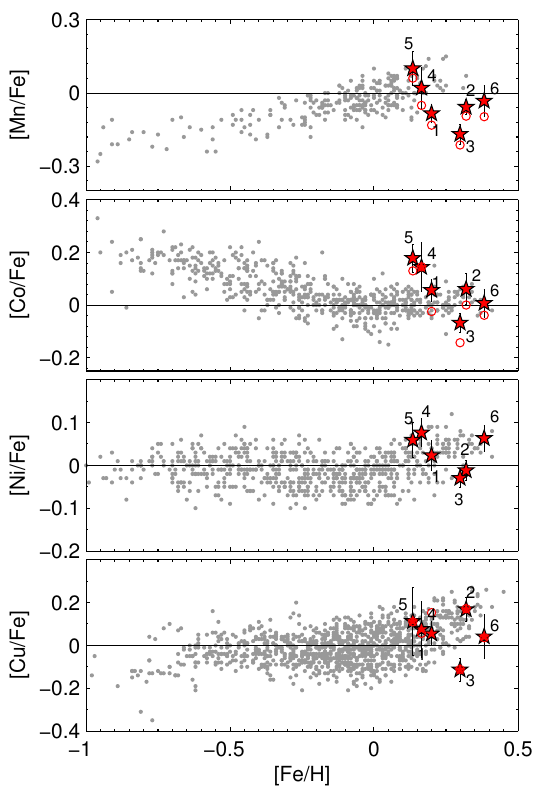}
\end{center}
\caption{Same as Fig.~3, but for the remaining Fe-peak elements. The reference samples were taken from 
\citet{Mishenina2015}, 
\citet{Battistini2015}, 
\citet{Bensby2014}, 
and \citet{DelgadoMena2017}. 
Red circles denote the values corrected for NLTE-effects, using data from \citet{Bergemann2008} for Mn,  \citet{Bergemann2010Co} for Co, and \citet{Caliskan2025} for Cu.}
\end{figure}
At their supersolar metallicities,  the [Mn/Fe] ratios of our stars start turning over from the Galactic trend of increasing Mn-abundance with 
increasing [Fe/H]  \citep{Mishenina2015} towards mostly sub-Solar ratios (top panel of Fig.~6). 
Here, we also show NLTE-corrected values from the line-by-line prescriptions of  \citet{Bergemann2008}, which lower the 
Mn abundance ratio by $\sim$0.05  dex. 

In the plot for [Co/Fe], one sees the same behavior as for the other Fe-peak elements in comparison with Galactic stars. 
As for other elements, Hip~23311 stands out with its elevated abundance, while Hip~9353 also displays an enhancement, albeit
with a larger error bar due to the use of only two lines with 
significantly differing EWs. 
For all measurements, we have applied the NLTE corrections from \citet{Bergemann2010Co}, on the order of 0.05--0.1 dex, 
with the corrected values indicated in Fig.~6. 

It is noticeable that the Mn/Fe, Co/Fe (and possibly Zr/Fe and V/Fe) ratios show downward trends with increasing [Fe/H], even if Hip~23311 was removed from the reasoning. This is unlikely a measurement artifact, since proper treatment of the hyperfine splitting was applied, and, despite Cu\,{\sc i}
also suffering strongly from that effect,  [Cu/Fe] is reasonably well behaved in our sample. 
This would suggest that this trend in our sample of metal-rich stars 
cannot be due to different dilutions of SNe ejecta with different
amounts of pure hydrogen. It rather indicates changing 
average yields with increasing metallicity in a chemically evolving system. 

Our work provides the first measurements of Cu in stars of this MG. 
To this end, three useful lines (at 5105, 5700, and 5782 \AA) were present in the spectra, with EWs typically on the order of  $\sim$100 \AA. 
To determine the [Cu/Fe] ratio, we resorted to synthesizing those lines. Their NLTE corrections are small \citep{Caliskan2025} and barely discernible in Fig.~6.  
At their slightly supersolar values, all stars (but one, Hip~22336)
nicely follow the trend outlined by Galactic disk stars \citep{Mishenina2011,DelgadoMena2017}. 
We note that the lower-Cu star Hip~22336 also shows the 
lowest [Mn/Fe] and [Co/Fe] ratios in our sample. 
As \citet{McWilliam2003} and \citet{McWilliam2005} showed, Cu-depletions are followed by deficiencies in [Mn/Fe] as well, 
which can be explained by low-metallicity SNe Ia and metallicity dependent (SNe Ia and SNe II) yields. However, those studies focused on the Sagittarius (Sgr) dwarf galaxy 
and neither our Cu- and Mn-depletions are as pronounced as in the lower-metallicity Sgr-stars.

Finally, Ni invariably traces iron as is expected given their lockstep production in regular Chandrasekhar-mass SNe Ia 
\citep[e.g.,][]{Woosley1995,Kobayashi2020}. 
{ \citet{Li2024a} used abundances from the APOGEE and GALAH surveys to argue for 
a ``low alpha thin disc'' HR~1614 subgroup 
to be enhanced in Na, Al, V, and Ni, while the enhancements we find here are mainly pertaining to those stars with a more likely thick disk origin.} 
\subsection{Heavy elements: Y, Zr, Ba, La, Eu}
In Fig.~7 we show the neutron-capture elements 
we measured in our stars. 
\begin{figure}[htb!]
\begin{center}
\includegraphics[angle=0,width=1\hsize]{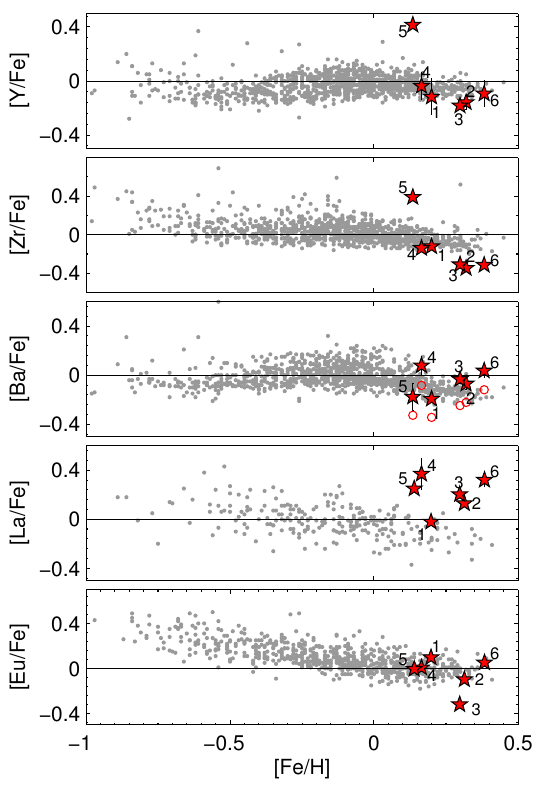}
\end{center}
\caption{Same as Fig.~3, but for neutron-capture elements. Reference data are from 
 \citet{Koch2002}, 
\citet{Bensby2014}, 
and \citet{DelgadoMena2017}. %
}
\end{figure}
The [Y/Fe] ratio of our stars, derived from spectral synthesis, is perfectly in line with the distribution in the disk, showcasing its origin in a standard, slow ($s$-) neutron capture process 
\citep[e.g.,][]{Busso1999,Tautvaisiene2021}. The usual contender for an outlying abundance is Hip~23311 with an overabundance of $\sim$0.4 dex.
This is also observed for Zr, where we find, however, that the three most metal-rich stars are depleted with respect to the remainder of 
metal-rich disk stars. 
We assessed NLTE-corrections on Zr by comparison with \citet{Zhao2016} and 
estimate departures of $\sim$0.1 dex only for the three warmer stars in our sample, with 
null corrections otherwise.

The Ba-abundances in our stars  are in very good agreement with the Ba-determination of  \citet{DeSilva2007} and with those in metal-rich disk stars. 
 Also the  EWs for star Hip~23311 agree excellently with their measurements with a  mean deviation of $-2.2\pm2.9$ m\AA~for the three lines in common. For reference, we indicate in Fig.~7
 Ba-values corrected for NLTE by using the line-by-line departures
 given by \citet{Gallagher2020}, albeit for the Sun. 
In turn, a broad range in [La/Fe] ratios is found from rather weak lines in our stars, with values lying above the disk trend by $\sim$0.4 dex.

Finally, the Eu abundances  align very well with the results from  \citet{DeSilva2007}, despite being based on only one line at 6645 \AA. 
NLTE-corrections are very small in the parameter space of T$_{\rm eff}$ and [Fe/H] of 
our stars \citep{Zhao2016} and we do not show them any further in Fig.~7. 
In particular, five of our six stars are fully consistent with the MW disk, while  Hip~22336 
is depleted by $\sim -0.3$ dex. This is particularly noteworthy, as this star was already shown to have systematically lower abundances of Mn, Co, and Cu
compared to the remainder of our sample. 

Our results indicate a high $hs/ls$-ratio in excess of 0.8 dex (taken here as the average of Ba and La relative to Y and Zr; listed in Table~A.2), which
normally is indicative for $s$-processing in metal-poor Asymptotic Giant Branch (AGB) stars
\citep[e.g.,][]{Busso1999}, due to the larger ratio of neutrons
per Fe-seed nucleus driving the synthesized elements to 
high mass numbers. 
However, in our sample we see high $hs/ls$-ratios at high metallicity, contrary to this expectation.  
This may be due to a higher, total neutron exposure in the polluting AGB stars. The usual deviant from this trend is Hip~23311 with a
$hs/ls$-ratio well below zero.
\subsection{Intrinsic abundance spreads}
\citet{Smith1983} used DDO photometry for 19 of the originally confirmed (26) metal-rich member stars of the HR 1614 MG and discovered anomalies 
in the sense of a large number of CN-rich stars. 
On the one hand, this finding cemented the original proposition of \citet{Eggen1978} of the physical association of the purported group's stars. 
However, \citet{Smith1983} argued from the then-available kinematics that those stars cover a broad range of orbits, with peri- and apocenter radii of 6.5--10.4 kpc,
and therefore must have formed in a region of larger extent than a typical 
(open) cluster. Reasoning from their CN-enhancements within the framework of a Galactic abundance
gradient in the old disk, they stated that spatial and temporal heterogeneities are likely to occur 
in the region within the disk where those stars have formed.  

In order to quantify the intrinsic abundance spread for each element, we follow the probabilistic approach of \citet[][in turn based on \citealt{Walker2006}]{PiattiKoch2018}. 
The resulting scatter $\sigma_0$, which also accounts for the measurement errors, is listed in the last column of Table~A.2 in the appendix. 
Here, the scatter in Fe-abundance amounts to 0.09 dex alone, which is significant at the 3$\sigma$-level. 
The highest spread of 0.32 dex  is found for Ti\,{\sc ii}, while it is lower for neutral Ti and thus could indicate 
a systematic effect in the measurements or NLTE effects. 
Similarly large spreads of 0.25 dex are seen for  Zr and C. The lowest value, in turn, is only found for the $\alpha$-elements S and Ca, 
which,  at a mere 0.02$\pm$0.02 dex each are consistent with a homogeneous origin. 

In conclusion, 21 out of the 27 abundances we measured show intrinsic spreads that are significantly above zero at least at the 2$\sigma$ level. 
This is in contrast to \citet{DeSilva2007} who found a very low scatter in most elements from their sample of 18 group members.
They stated an intrinsic scatter around 0.01--0.03 dex (at the 80\% C.L.) for most of the studied elements. Coupled with the 
 coetaneous ages of the group stated by \citet{DeSilva2007}, this chemical homogeneity was taken as evidence for the MG originating 
 in an unperturbed, dispersed star-formation event. 
Our finding of significant intrinsic abundance spreads also bolsters previous results that 
the HR~1614 MG is not a dissolved open cluster \citep{Kushniruk2020}: 
in their sample of 26 open clusters, \citet{Sinha2024} determined 
very low spreads below $<$0.02 dex (at the 3$\sigma$-level) for most of the (20) measured elements from their APOGEE cluster sample, even  though a different metric from ours was used. 
Exceptions to this lack of a dispersion were mainly found for the weak-lined neutron-capture elements, resulting in larger spreads of $\sim$0.2 dex.

The large abundance spreads seen in our analysis rather argue in favour of dynamical mixing events bringing in together stars from various environments,
in particular, as MW stars tend towards chemical homogeneity to within $\sim$0.02 dex at a given Galactocentric radius \citep{Ness2022}. 
{ 
As we will also reason in Sect.~6, at least part of our sample has a thick disk origin. }
\section{Ages}\label{sec:Ages}
While \citet{Feltzing2000} obtained a single age of $\sim$2 Gyr with the possibility of some spread for the HR~1614 MG as an ensemble, 
\citet{Kushniruk2020} derived ages from an unprecedentedly large number of 
$\mathcal{O}(10^4)$ stars using Gaia photometry and photometric metallicities from the SkyMapper survey 
\citep{Keller2007}, but no comments on individual stars were made. 
Albeit our smaller sample size, we are in the fortunate situation of having precise and accurate abundance measurements 
that we use as an input for fitting isochrones in an effort to derive individual stellar ages of our targeted group members. 

In practice, we estimated ages  via fitting  the suite of Dartmouth isochrones\footnote{These do not account for stellar rotation as our stars, at M$\la$1.3 M$_{\odot}$ 
\citep{DeSilva2007}, are likely not strongly affected by this effect. Switching to the PARSEC isochrones that include rotation above $\sim$1~M$_{\odot}$ 
\citep{Nguyen2022}
has little impact on the ages derived here.} 
\citep{Dotter2008},  tailored to each star's individual metallicity,  
 to the targets' location in a color-magnitude diagram from Gaia photometry \citep[][see Fig.~8]{GaiaDR3}. 
To this end, we  employed 
 distances and extinctions from \citet{BailerJones2018} and \citet{Queiroz2018}, 
and the extinction law of \citet{Yang2023}. 
\begin{figure}[htb!]
\begin{center}
\includegraphics[angle=0,width=1\hsize]{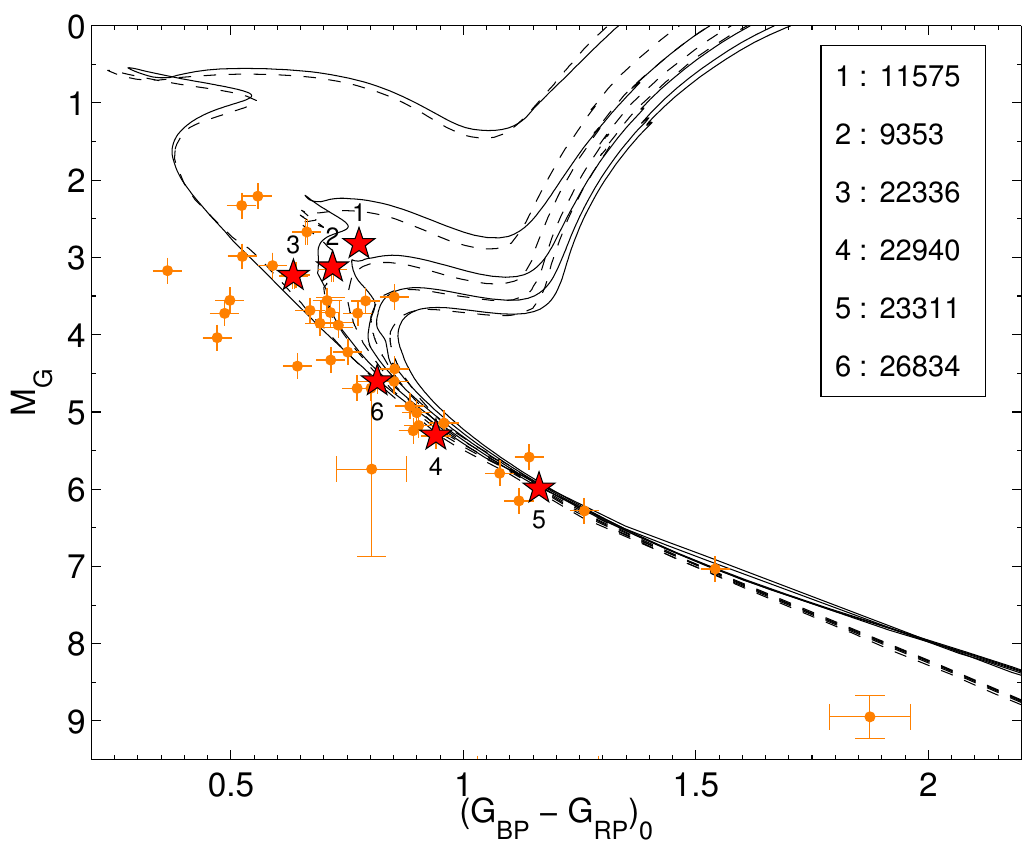}
\end{center}
\caption{Color-magnitude diagram of our target stars (red symbols) and those of \citet[][orange points]{Feltzing2000} with the set of Dartmouth- 
(\citealt{Dotter2008}; solid lines) and the rotating PARSEC-isochrones 
(\citealt{Nguyen2022}; dashed lines) used for age estimates. While each star was fit with an isochrone of its individual metallicity, 
we show here an exemplary set of [M/H]=0.2 dex for ages of 1 to 9 Gyr in steps of 2 Gyr. As before, number labels refer to Hip-numbers.}
\end{figure}

    As Fig.~8 indicates, three stars of our sample lie below the age-sensitive main-sequence turn-off and the isochrone fits merely return the oldest
age used in the grid. For the other MG member stars, we obtain ages of 3.0$\pm$0.8 Gyr (Hip~11575), 2.8$\pm$0.4 Gyr (Hip~9353), and
4.4$\pm$0.7 Gyr (Hip~22336), respectively. 
The stated age errors were  based on the photometric and parallax errors provided by the \citet{GaiaDR3}, and by accounting for a metallicity uncertainty of $\pm$0.1 dex. 
 This leads to relative age errors on the order of 15--30\% for the three reliable measurements.\footnote{We also employed the  [Y/Mg] ratio as a a chemical clock, using the calibration of \citet{Casali2020} to estimate  ages of $\sim$1.4$\pm$0.8 dex for
Hip~11575, 9353, and 22336, and 2.4$\pm$1.0 Gyr for Hip~26384, while for the other two, no physically meaningful values could be determined.}

The range of possible ages of the HR~1614 MG we find emphasizes the possibility that members of this  group can indeed be older and diverse stars, 
which is in fact in line with \citet{Hufnagel1994}, who found from chromospheric activity indicators 
that their analysed stars in this MG are  older than 3 Gyr.
Previous estimates from canonical isochrone fitting, as we have employed here, have placed the age of this moving group at a slightly younger age of around 2 Gyr \citep{Feltzing2000}, 
although \citet{Kushniruk2020} assert that a broad age range with prominent peaks around $\sim$2 and $\sim$8 Gyr indicate stars of a multifarious disk origin within the MG.

We therefore summon up the 
same exercise as above  for the {\em bona fide} sample from \citet{Feltzing2000}, bearing in mind that their metallicity scale from Str\"omgren 
photometry and, for a handful stars, [Fe/H] from spectroscopic measurements from the literature,  
will be different from ours, which poses a possible cause for systematic offsets. As before, we obtained those stars' photometry and parallaxes from the 
\citet{GaiaDR3}. The resulting age distribution is shown in Fig.~9. 
\begin{figure}[htb!]
\begin{center}
\includegraphics[angle=0,width=1\hsize]{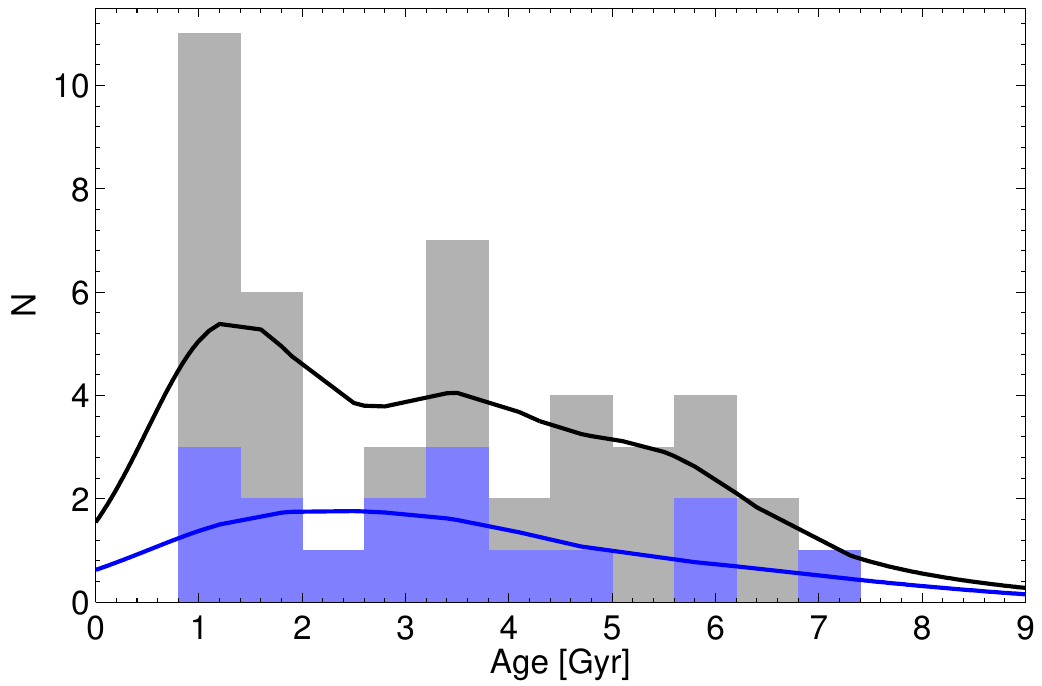}
\end{center}
\caption{Histogram of ages derived from the sample of \citet{Feltzing2000}. The entire sample is shown as gray-shaded, while stars on the upper, more age-sensitive
CMD are highlighted in blue. Solid lines show the respective error-weighted distributions.}
\end{figure}

As a considerable fraction of the stars lie below the respective turn-offs on the rather age-insensitive main sequence, we also 
constructed the same histogram for stars brighter than the (arbitrary) limit of M$_G < 4$ mag.
It becomes immediately obvious, as also implied by the color-magnitude distribution of the sample,  that the stars in this group 
span a large range of ages. The strong peak at the youngest ages (most strongly visible in the full distribution shown in gray in Fig.~9) 
can be attributed to the limit of our fitted isochrone grid and 
therefore represents an artifact. Nonetheless, each distribution shown in Fig.~9 displays clear maxima of up to 
2 and at $\sim$3.5 Gyr.
Further peaks at 5 Gyr and above $\sim$6.5 Gyr are also discernible, 
though less pronounced.

Most likely is a coincidence of this dynamical group with 
the 
interaction with the Galactic bar, which 
we explore in greater detail in the next Section~6. 
The bar can have already formed early on 
with a steep cessation in star formation as, e.g., 
found from studies of its supermetal-rich populations
(\citealt{Nepal2024}; see also \citealt{Sanders2024}).
A too young age of the bar itself is dynamically not feasible or likely
 \citep[e.g.,][]{Debattista2019,Anderson2024}.
\section{Dynamics}
Radial velocities were derived using a cross-correlation of the spectra against a synthetic spectrum using IRAF's {\em fxcor} task \citep{TonryDavis1979}, 
where we restricted our measurements to the less-blended wavelength region of 5700--6400 \AA. 
The resulting velocity uncertainties  are on the order of 0.1 km\,s$^{-1}$.
Our derived velocities are in good agreement with those from the literature (\citealt{DeSilva2007}; \citealt{Feltzing2000}, and references therein), with 
an average difference of $-0.2$ km\,s$^{-1}$. Slightly larger deviations are only found for Hip\,23311 and Hip\,26834, for both of which \citet{Feltzing2000}
reported velocities larger by 6--7  km\,s$^{-1}$  than our and De Silva's (2007) values. 
Furthermore, proper motions were taken from the third data release of Gaia \citep{GaiaDR3}, 
along with distances, extinctions and reddening corrections from \citet{BailerJones2018} and \citet{Queiroz2018}.
The observed and derived properties of our sample  are detailed in Table~B.1 in the appendix. 
\subsection{Orbit integration}
With a bar resonance being a strong contender for having 
{ stirred up 
stars into their dynamical coherence, 
} particular care has to be applied to the potential 
of that particular MW component.

{Thus, 
we used the \texttt{AGAMA} package \citep{Vasiliev2019_AGAMA} together with the MW potential model of \citet{Hunter2024}. This model provides a self-consistent non-axisymmetric description of the inner Galaxy. The potential includes the central black hole, the nuclear star cluster, the nuclear stellar disk, and the dark halo in an axisymmetric multipole component, while the bar, thin and thick stellar disks, and the gas disks are represented by a triaxial 
\texttt{CylSpline}\footnote{ 
Here, \texttt{CylSpline} refers to AGAMA's cylindrical-spline representation of the potential, in which the radial and vertical dependence is tabulated on a cylindrical grid and the azimuthal structure is represented by Fourier harmonics.} component. }

{ The bar rotates clockwise
in the Galactocentric $(x,y)$ plane, i.e. in the same sense as Galactic disk rotation in the adopted AGAMA convention; throughout this work we quote the absolute value of the pattern speed, with a pattern speed of $\Omega_{\rm bar}=37\,{\rm km\,s^{-1}\,kpc^{-1}}$ and is oriented at an angle of $27^\circ$ with respect to the Sun--Galactic-centre line.}

{ For the analysis presented here we used the rotating version of the model without spiral arms. 
For comparison, we also tested the corresponding rotating model including spiral arms. 
The optional spiral component in the Hunter et al. model is implemented as a separate, more slowly rotating \texttt{CylSpline} perturbation. 
It is constructed from two phase-shifted $m=2$ logarithmic spiral sets and represented with azimuthal harmonics up to $m_{\max}=8$.
For the six stars, the inclusion of the spiral component only has an insignificant impact on the derived parameters: the
pericentres change by at most $0.032$\,kpc, apocentres by $0.072$\,kpc, eccentricities by $0.007$, and $Z_{\rm max}$ by $0.003$\,kpc. These correspond to relative changes below $0.9\%$ for the orbital radii and less than $2.5\%$ for $Z_{\rm max}$.
}

{For each star, the observed phase-space coordinates were transformed into a Galactocentric frame assuming $R_0=8.179$\,kpc, $z_\odot=20.8$\,pc, and a solar velocity of $(U,V,W)_\odot=(11.1,245.0,7.25)\,{\rm km\,s^{-1}}$. The resulting Cartesian coordinates and velocities were then rotated into the bar frame before orbit integration. Orbits were integrated for 5\,Gyr using the rotating Hunter et al. potential and sampled with $10^5$ trajectory points. From the resulting trajectories we derived pericentres, apocentres, eccentricities, maximum vertical excursions, angular momenta, and energies. In addition, we computed the fundamental orbital frequencies in both the inertial and the bar-rotating frame.}

{Uncertainties in the derived orbital parameters were quantified with a Monte Carlo sampling procedure. For each star, we constructed a covariance matrix from the measured radial velocity, distance, and proper motions ($\mu_\alpha$, $\mu_\delta$), including the variances of the individual quantities and the covariance between the two proper-motion components. We then drew 1000 Monte Carlo realisations per star. For each realisation, the full coordinate transformation, rotation into the bar frame, orbit integration, and frequency determination were repeated. The reported orbital quantities are the median values of the Monte Carlo
realisations, while the 16th and 84th percentiles are used as a robust $68\%$ uncertainty interval.}

\subsection{Orbital parameters}
The estimated orbital parameters of our stars and their uncertainties are listed in Table~B.1, and illustrations of  
the orbits in the restframe of the bar are shown in Fig.~B.1, each in the
appendix.

{All six orbits remain confined close to the Galactic mid-plane in the rotating potential. The maximum vertical excursions span
$Z_{\rm max}=0.036$--$0.238$\,kpc. \object{Hip 22336} reaches the largest height above the plane, $Z_{\rm max}=0.238$\,kpc, whereas \object{Hip 22940} and \object{Hip 23311} remain very close to the plane, with $Z_{\rm max}=0.036$ and $0.053$\,kpc, respectively.}

{The overall structure of the orbits is remarkably homogeneous. Apocentres cluster tightly around $R_{\rm Apo}=8.53$\,kpc, with a full range of only $0.17$\,kpc, while pericentres are found around $R_{\rm Peri}=3.64$\,kpc, spanning $0.36$\,kpc. This yields eccentricities in the narrow interval $e\simeq0.39$--$0.44$ for all six stars. The most eccentric orbit is that of \object{Hip 22336}, with $e=0.436$, which is also the star with the smallest pericentre, $R_{\rm Peri}=3.37$\,kpc.}
The orbital periods about the Z-axis are also nearly identical, at $P_z\simeq0.16$–$0.17$\,Gyr. We note that this coincides with the period of the bar for our assumed pattern speed of $\Omega_{\rm bar}=37\,{\rm km\,s^{-1}\,kpc^{-1}}$,
so that in the rotating frame of the bar the Jacobi integral ($E_{\rm J}$, also listed in Tab.~B.1) is approximately conserved and all six stars occupy a very narrow range in $E_{\rm J}$.

{The angular momenta and present-day azimuthal velocities are also very similar across the sample. In the sign convention of the present integration, the prograde orbits have positive angular momenta, with $|L_z|=1.41$--$1.55\times10^3$\,kpc\,km\,s$^{-1}$. 
\object{Hip 22336} is the kinematically warmest object in the sample, with the lowest $|L_z|$, the lowest $|V_\phi|$, the largest vertical excursion, and the largest eccentricity.}

\begin{figure}[htb!]
\begin{center}
\includegraphics[angle=0,width=1\hsize]{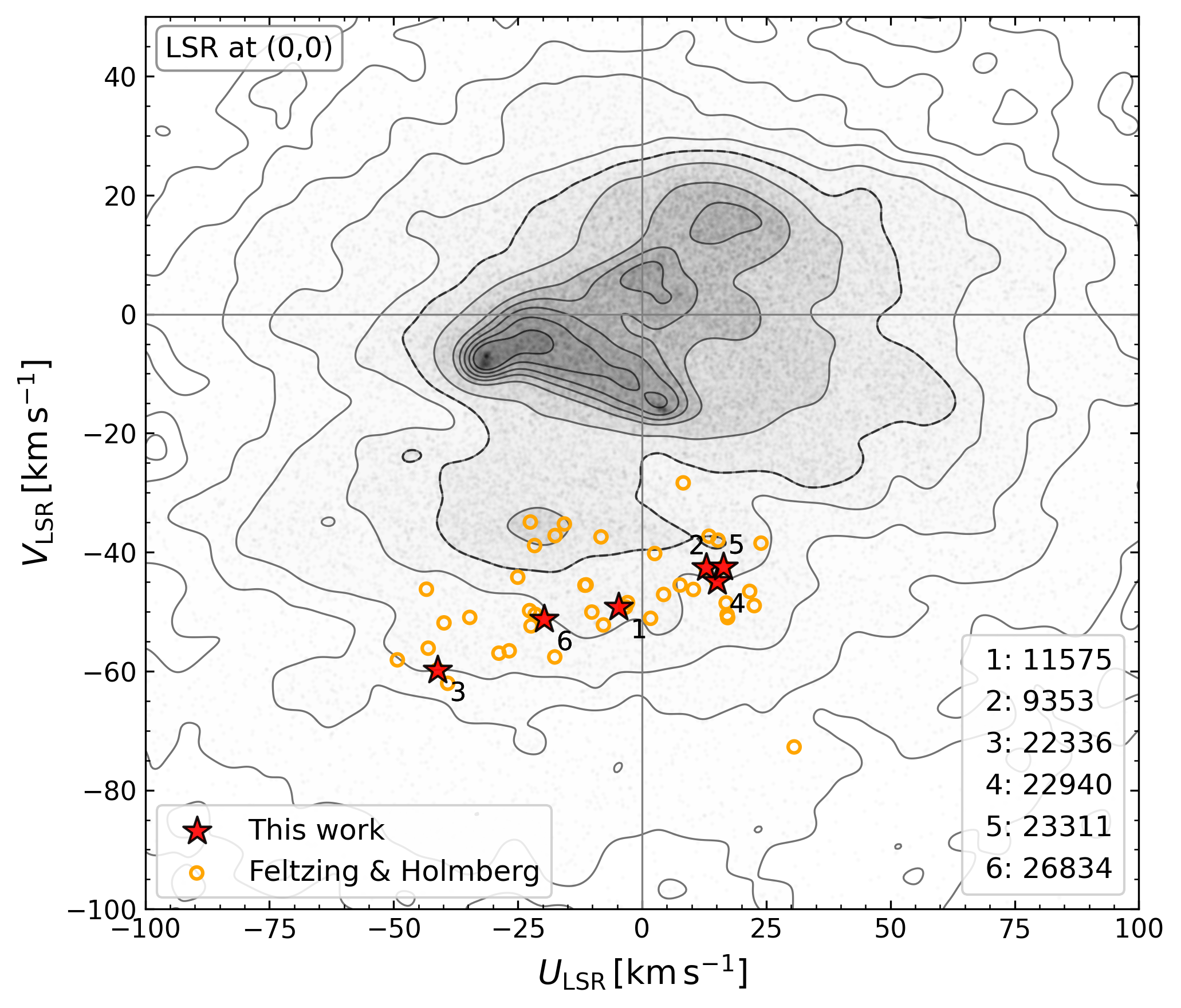}
\end{center}
\caption{Velocity distribution in the solar neighborhood showing the GALAH field-star density (grey shading) with kernel–density contours (black) in the $(U_{\rm LSR},V_{\rm LSR})$ plane. Red stars mark our MG members; open orange circles show the comparison sample from \citet{Feltzing2000}. The crosshair indicates the Local Standard of Rest at $(0,0)$. Numbers next to the red symbols correspond to the Hip-identifiers listed in the legend.}
\label{fig:UV_plot}
\end{figure}

\begin{figure}[htb!]
\begin{center}
\includegraphics[angle=0,width=1\hsize]{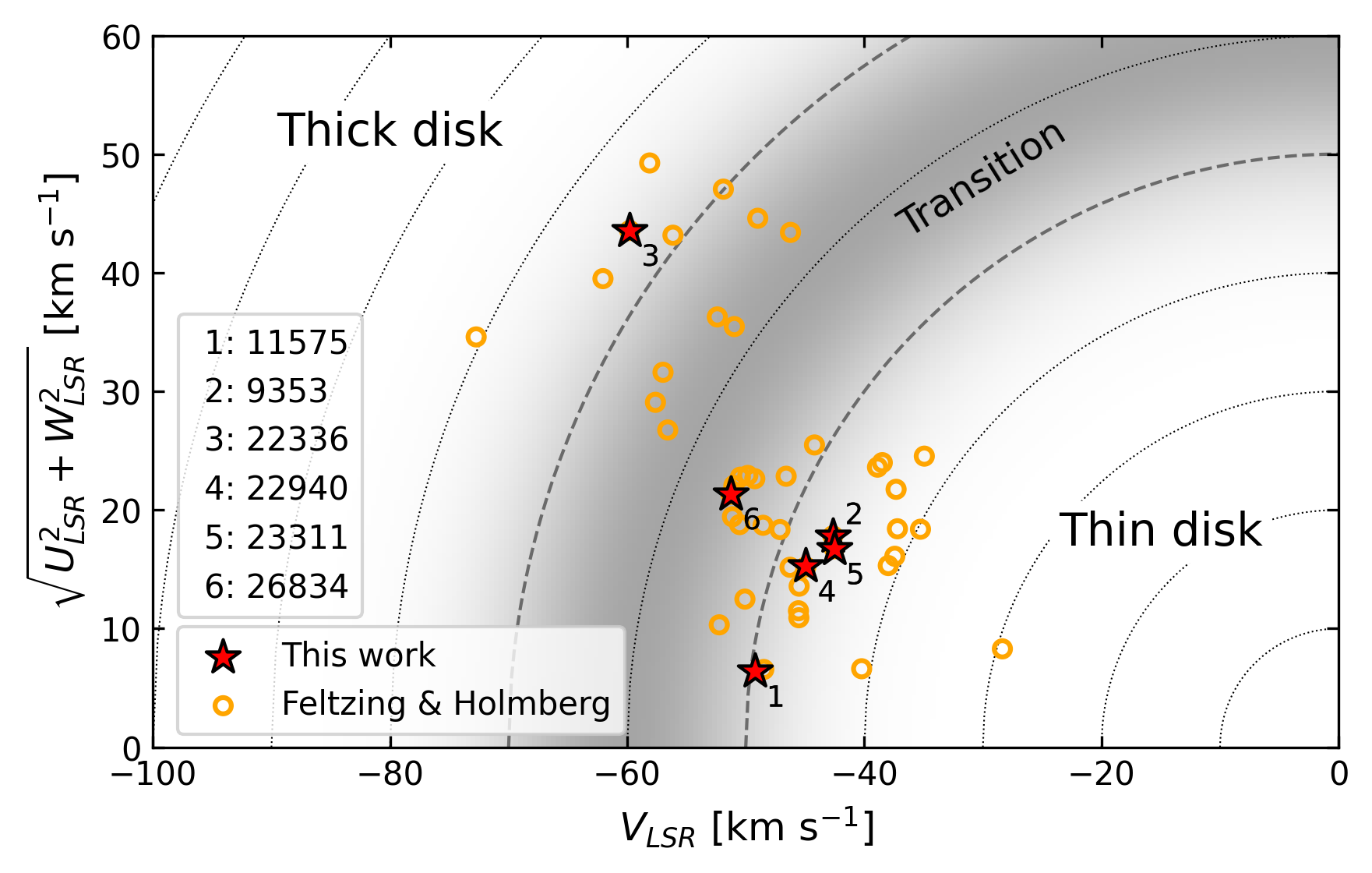}
\end{center}
\caption{Toomre diagram for HR\,1614 MG members. Red stars mark stars analyzed in this work (numbers indicate Hip-identifiers); orange open circles show members from \citet{Feltzing2000}. Dotted curves trace illustrative loci of constant total space velocity with respect to the LSR. The shaded background with dashed guides only schematically indicates the thin–disk, transition, and thick–disk regimes commonly used in Toomre space \citep[e.g.,][]{Bensby2003}.}
\label{Fig:Toomre_diagram}
\end{figure}

In the local velocity plane (Fig.~\ref{fig:UV_plot}), the MG members also form a compact, though tilted sequence at $V_{\rm LSR}\sim -50$\,km\,s$^{-1}$ that follows and slightly extends the  group as selected by \citet{Feltzing2000}.
The Toomre diagram (Fig.~\ref{Fig:Toomre_diagram}) reinforces the formation picture of the HR~1614 MG. For each star this shows its Galactic rotational component ($V_{\rm LSR}$) against the vertical and radial deviations thereof, 
i.e., 
$T\equiv\sqrt{U_{\rm LSR}^2+W_{\rm LSR}^2}$. Five of our target stars lie in the thin–disk locus with modest $T$ and mildly lagging $V_{\rm LSR}$. \object{Hip 22336} is a clear kinematic outlier, placing it on the transition band toward the thick disk. This behavior is consistent with its larger $Z_{\rm max}$, lower $L_z$, and higher $|U_{\rm LSR}|$. Overall, the distribution of our analyzed stars overlaps the locus of literature \object{HR 1614} MG members from \citet{Feltzing2000}, yet \object{Hip~22336} sits near the upper edge of that cloud
together with a subgroup of another dozen of stars with rather thick 
disk kinematics.

Thus, about two thirds of the MG stars considered here are compatible with 
thin-disk stars, while the remainder, and markedly Hip~22336, is found at a higher deviation from the circular V$_{\rm LSR}$-component (i.e., high $T$), and located toward thick-disk stars. 
{ We thus ascertain a dynamical admixture from the thick disk during the 
resonant formation 
of this MG}, as also indicated by the advanced age of those stars. Additionally, the elevated  Z$_{\rm max}$ and eccentricity of such stars, in particular Hip~22336, are an indicator of a fraction of the HR~1614-group stars coming from the thick disk. The star with the strongest thick-disk like chemical evidence, however, is kinematically closer to the thin disk.
This is in line with the existence of kinematically thick
disk stars among the chemically thin disk population, and vice versa.  As argued from the chemical analysis above (Sect.~4.3) this makes star Hip~22336 (with thin-disk abundances and thick-disk dynamics) a common representative of the chemodynamic crucible of this MG (see also \citealt{McWilliam2016}; their Fig.~2).
{ Finally, we note that the inclusion of the bar in the potential on top of a 
regular, axisymmetric potential \citep[e.g.,][]{McMillan2017}
results in a decrease of the mean Galactocentric radii of the stars (as detailed in Appendix~B and Table~B.2), highlighting 
the importance of the bar for inducing 
significant orbital changes.}
\subsection{Orbital resonances}
\begin{figure}
    \centering
  \includegraphics[width=1\linewidth]{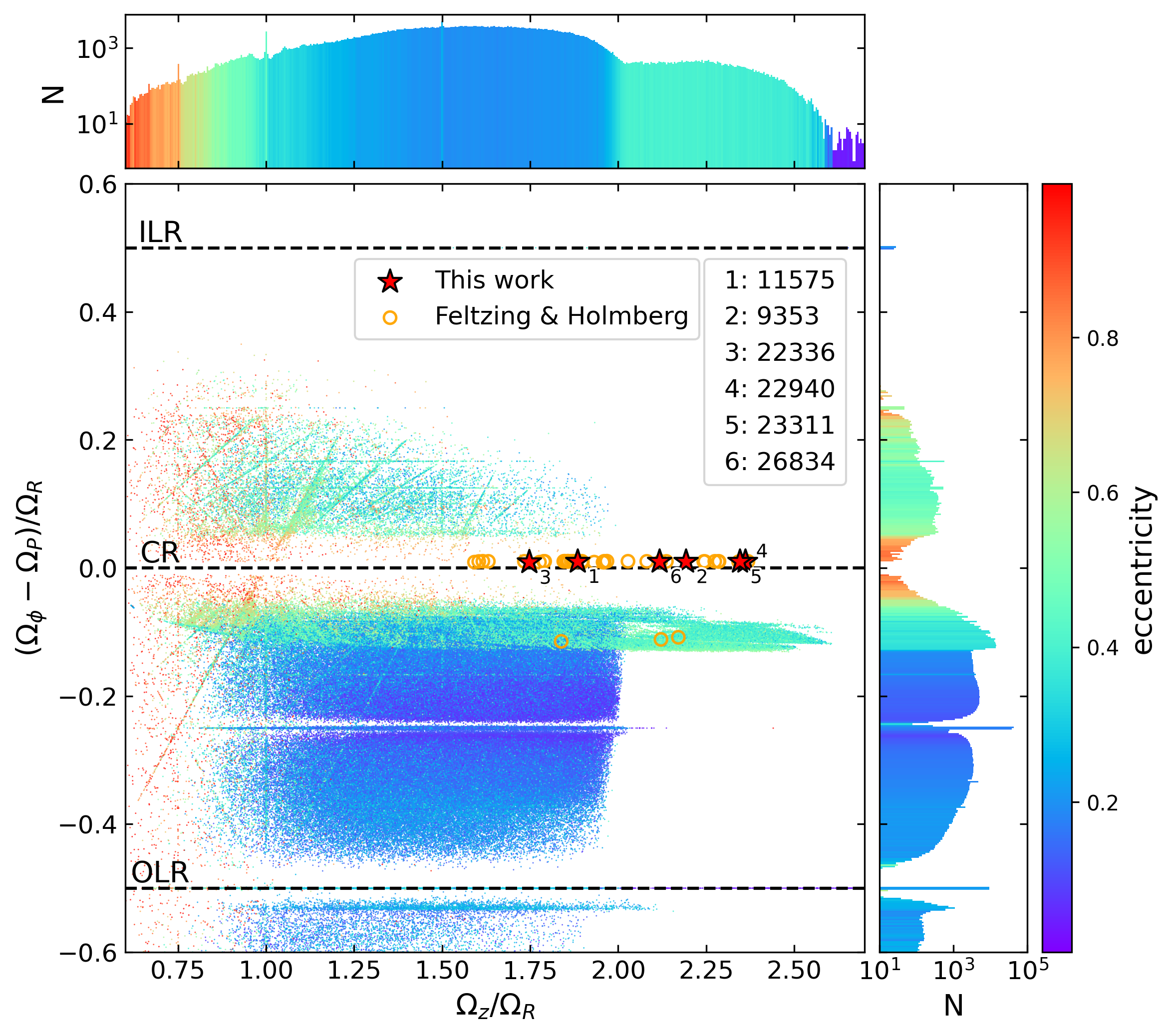}
    \caption{Frequency-ratio map for the \object{HR 1614} MG relative to a rigidly rotating non–axisymmetric pattern. The main panel shows $(\Omega_\phi-\Omega_{\rm p})/\Omega_R$ vs. $\Omega_z/\Omega_R$; background points (color–coded by eccentricity) depict a comparison disk sample, with the top and right panels giving the corresponding 1D distributions. Horizontal dashed lines indicate the resonant loci for an $m=2$ perturbation: corotation (CR; $(\Omega_\phi-\Omega_{\rm p})/\Omega_R=0$), inner Lindblad resonance (ILR; $+1/2$), and outer Lindblad resonance (OLR; $-1/2$). 
    Red symbols mark the six stars analysed in this work (numbers label \textit{Hipparcos} IDs); orange open circles show literature 
    candidates from \citet{Feltzing2000}. 
    { The three orange points below the CR line at $y\simeq-0.09$ form a higher-angular-momentum tail of the literature comparison sample.}
    }
    \label{fig:frequencies}
\end{figure}

Figure~\ref{fig:frequencies} shows a frequency map for our MG stars against a large comparison sample of $\sim$4.6$\times10^5$ disk stars
from \textit{GALAH} DR4 \citep{Buder2025}. We computed the three fundamental orbital frequencies, $\Omega_R$, $\Omega_\phi$, and $\Omega_z$, which characterize the radial epicycle, the mean azimuthal rotation, and the vertical oscillation of an orbit \citep[e.g.][]{BinneyTremaine2008}. 
In the inertial frame our adopted MW mass model is time dependent, whereas in the frame corotating with the bar at constant pattern speed $\Omega_p$ it is time independent (i.e., steady). 
{ In either case, 
resonant lines appear as straight loci in the $(\Omega_z/\Omega_R,\ (\Omega_\phi-\Omega_p)/\Omega_R)$ plane} \citep{Valluri2010, Valluri2016, Bovy2026}. Frequencies were measured from the integrated phase–space time series with the \texttt{naif} Python package, which implements the Numerical Analysis of Fundamental Frequencies (NAFF) algorithm \citep{Laskar1993, Valluri1998, BeraldoSilva2023}. This spectral approach robustly recovers the dominant quasi–periodic components and is well suited to highlight bar resonances as straight loci in frequency–ratio space \citep{Monari2019}. 

We integrated the orbits in our adopted MW potential and determined the inertial-frame frequencies $(\Omega_\phi, \Omega_R, \Omega_z)_{\rm ini}$. 
For a rotating bar with pattern speed $\Omega_p$, we transformed these to the bar-frame by subtracting the pattern rotation from the azimuthal angle and the corresponding velocity field, which then allowed us to 
re-measure the bar-frame frequencies \((\Omega_\phi,\Omega_R,\Omega_z)_{\rm bar}\). 
As a consistency check we verified that $\Omega_{\phi,{\rm ini}}-\Omega_p \simeq \Omega_{\phi,{\rm bar}}$ to within numerical tolerance. In Fig.~12,  we show the dimensionless ratios $x \equiv \Omega_z/\Omega_R$ and $y \equiv (\Omega_\phi-\Omega_p)/\Omega_R$, which compactly encode in–plane and vertical frequency information 
\citep{BinneyTremaine2008, Monari2019}. {For the six stars analysed in this work, the plotted quantities are based on the same Monte Carlo realisations used for the orbital parameters.}

In a rotating non-axisymmetric potential the general resonance condition is $m \big(\Omega_\phi-\Omega_p\big) = \ell \,\Omega_R + n\,\Omega_z$, with integers $(m,\ell,n)$. For a bisymmetric bar $(m=2)$,  
$(n=0)$, in–plane resonances  appear at corotation ($y=0$) and as the ILR and OLR, respectively ($y=\pm\tfrac{1}{2}$).
Vertical families $(n\neq 0)$, in turn, map to straight lines $y=(n/m)\,x-\ell/m$ and generate the fan of thin slanted ridges in the background distribution of Fig.~\ref{fig:frequencies}, where the vertical oscillation frequency is commensurable with the combination of radial and azimuthal motion. In our fully three-dimensional non-axisymmetric MW potential several low-order $(\ell,n)$ families overlap in frequency space, so individual ridges are blends rather than pure resonances; we therefore only refer to them generically as vertical $(n\neq 0)$ families. 

The concentration of points along the CR line and the clarity of the ILR/OLR bands are a direct consequence of using a dynamically consistent potential and a pattern speed matched to it: with a mismatched $\Omega_p$, the same stars would shear vertically (changing $y$) and the resonant ridges would blur \citep{Dehnen2000, Monari2019}. That the resonant structure emerges cleanly is therefore a useful validation of our choice of both potential  and frequency-extraction method.

For context, we overplot in Fig.~12 a literature comparison set drawn from \textit{GALAH} DR4 \citep{Buder2025} following the best-practice recommendations of DR4 with $\text{S/N}>30$ 
and supplemented by conservative cuts (\texttt{vsini}$<15\,\text{km s}^{-1}$, \texttt{e\textunderscore rv\textunderscore comp\textunderscore1}$<1\,\text{km s}^{-1}$). 

{Our six MG stars 
and all but three of the literature stars from \citet{Feltzing2000} 
occupy a compact region in frequency space. They span approximately $1.75\lesssim\Omega_z/\Omega_R\lesssim2.36$ and lie very close to the CR line, with $(\Omega_\phi-\Omega_p)/\Omega_R\simeq0$. 
We therefore interpret them as stars on  CR
rather than as members of other, distinct resonant families. 
The spread in $\Omega_z/\Omega_R$ reflects the different vertical orbital structure of the stars, consistent with the small but measurable differences in $Z_{\rm max}$ and eccentricity discussed above.} We also note the the small values of the stars' frequency drift\footnote{
$\Delta\Omega=\max_{i=1,2,3}\left|\frac{\Omega_i(T_2)\,-\,\Omega_i(T_1)}{\Omega_i(T_1)}\right|$, where the three  frequencies 
are evaluated in two separate parts  ($T_1$, $T_2$) over the orbits
\citep{Valluri2010}.}, $\log{\Delta\Omega}$ at CR (see also Table B.~1), which indicates no significant chaotic behavior in their orbits.

{ In contrast, three \citet{Feltzing2000} candidates below the CR line are offset to $y\simeq-0.09$. In the same calculation they have larger absolute angular momenta, $|L_z|\simeq1.60$--$1.66\times10^3$\,kpc\,km\,s$^{-1}$, and larger present-day azimuthal velocities, $|V_\phi|\simeq197$--$204$\,km\,s$^{-1}$, than the remainder of the sample. 
We regard them as a higher-angular-momentum extension of the original literature selection rather than as part of the very compact CR clump traced by our sample.}
{ Finally, we note that the  above CR coordinate $y$ 
changes by less than $10^{-4}$ upon including the spiral-arm potential, so our identification of the CR resonance remains unaffected.}

The broad distribution of disk stars outlines well-known resonant ridges associated with the Galactic bar. Similar structures have been linked to the Hercules stream and other local MGs in both models and data \citep{Dehnen1998, Dehnen2000, Monari2017, Monari2019}. The alignment of our stars with the CR ridge
then supports the interpretation that they belong to a resonantly influenced disk population. 
\section{Discussion and conclusions}
Even though the entire HR~1614 MG contains a
 large number of $\mathcal{O}(10^4)$ 
candidates \citep{Kushniruk2020}, the six members studied here at high resolution 
and in great chemical detail offer a valuable insight into 
the origin of those stars. In particular, 
by accounting for a realistic potential of the rotating bar
and from the ensuing frequency analysis, 
{ we established that the stars' common motion originated from a CR 
resonance with the MW bar}.
Overall, our results are in very good agreement with the chemical abundance study of \citet{DeSilva2007}.
Notably, Hip~23311 is chemically in line with a thick disk heritage, while Hip~22336, and possibly Hip~26834, are more closely kinematic members of the thick disk than the thin disk.

Numerous mechanisms can perturb any pre-existing population such as the MW disk(s). 
For instance, 
\citet{Minchev2009} investigated a merging dwarf galaxy, which can induce stream-like perturbations in the form of a 
``ringing'' disk. 
The ensuing phase-space waves by such mergers 
rather constitute dispersed star formation events
that would lead to chemical element spreads, leaving 
a chemical signature dissimilar from disrupted clusters
\citep[e.g.,][]{Kushniruk2019}. 
Crossings through the transient spiral arms also play a role in the formation of stream-like structures 
\citep{Quillen2018a,Quillen2018b}, 
although { this interpretation is
ruled out by} the older HR~1614 MG.

The Galactic bar is responsible for the formation of many observed, coherent dynamical features 
\citep[e.g.,][]{Dehnen2000,Trick2021,Wheeler2022}. 
{ In particular, the 
bar is able to trap stars
of any ages, so that any dynamical influence is not restricted to stars younger than the bar. }
{ \citet{Li2024b,Li2024a} performed a chemodynamical study from archival APOGEE and GALAH data of the Hercules stream, of which 
the HR~1614 MG is a purported subgroup, adding further insight into the complex effects of resonant disk-stirring.} 

The six stars studied here can only graze the surface of 
understanding the origin of the large dynamical groups.
Therefore it is imperative to also exploit 
the ongoing or upcoming large spectroscopic surveys to sample their grand abundance space, such as
offered by the 4MOST, MOONS, WEAVE, or SDSS-V 
Skymapper campaigns
\citep{Cirasuolo2011,deJong2012,Kollmeier2017,Bensby2019,Gonzalez2020}.
\begin{acknowledgements}
We are grateful to J. Bovy, E. Caffau, W. Dehnen, and S. Reffert for helpful discussions.
This research made use of atomic data from the INSPECT database (\url{www.inspect-stars.com}).
 \end{acknowledgements}
\bibliographystyle{aa} 
\bibliography{ms} 
\appendix
\section{Abundance errors and results}
%
%
%
\begin{table*}[h!]
\renewcommand\thetable{A.1}
\caption{Systematic errors, exemplary for star Hip 11575}
\centering
\begin{tabular}{ccccccc}
\hline\hline       
 & T$_{\rm eff}$ & log\,$g$ & [Fe/H] & v$_{\rm mic}$ & &  \\
\rb{Species} & $\pm\,$100 K & $\pm\,$0.10 dex & $\pm\,$0.10 dex & $\pm\,$0.10 km\,s$^{-1}$ & \rb{{\sc AODFNEW}} & \rb{$\sigma_{\rm Tot}$} \\
\hline       
C I   &   $\mp$0.08 & $\pm$0.04 &   $<$0.01 &   $<$0.01 & $+$0.05 & 0.09\\
O I   &   $\mp$0.11 & $\pm$0.03 & $\pm$0.01 & $\mp$0.01 & $+$0.10 & 0.11\\
Na I  &   $\pm$0.05 & $\mp$0.02 &   $<$0.01 & $\mp$0.01 & $+$0.02 & 0.05\\
Mg I  &   $\pm$0.04 & $\mp$0.01 &   $<$0.01 & $\mp$0.01 & $<$0.01 & 0.04\\
Al I  &   $\pm$0.04 &   $<$0.01 &   $<$0.01 & $\mp$0.01 & $-$0.01 & 0.04\\
Si I  &   $\pm$0.02 &   $<$0.01 &   $<$0.01 & $\mp$0.01 & $+$0.02 & 0.03\\
K I   &   $\pm$0.08 & $\mp$0.04 & $\pm$0.01 & $\mp$0.03 & $+$0.04 & 0.09\\
Ca I  &   $\pm$0.06 & $\mp$0.02 &   $<$0.01  & $\mp$0.03 & $+$0.01 & 0.07\\
Sc II &   $\mp$0.01 & $\pm$0.04 & $\pm$0.02 & $\mp$0.03 & $+$0.11 & 0.06\\
Ti I  &   $\pm$0.09 &   $<$0.01 &   $<$0.01  & $\mp$0.01 & $-$0.02 & 0.09\\
Ti II &   $\mp$0.01 & $\pm$0.04 & $\pm$0.02 & $\mp$0.05 & $+$0.10 & 0.07\\
V I   &   $\pm$0.10 &   $<$0.01 &   $<$0.01 & $\mp$0.01 & $-$0.02 & 0.10\\
Cr I  &   $\pm$0.08 & $\mp$0.01 &   $<$0.01 & $\mp$0.03 & $-$0.01 & 0.08\\
Mn I  &   $\pm$0.09 & $\mp$0.01 & $\mp$0.01 & $\mp$0.06 & $<$0.01 & 0.11\\
Fe I  &   $\pm$0.07 & $\mp$0.01 &   $<$0.01 & $\mp$0.04 & $+$0.01 & 0.08\\
Fe II &   $\mp$0.03 & $\pm$0.04 & $\pm$0.02 & $\mp$0.04 & $+$0.11 & 0.08\\
Co I  &   $\pm$0.08 &   $<$0.01 &   $<$0.01 & $\mp$0.01 & $-$0.01 & 0.08\\
Ni I  &   $\pm$0.07 &   $<$0.01 &   $<$0.01 & $\mp$0.03 & $+$0.01 & 0.07\\
Cu I  &   $\pm$0.10 & $\mp$0.01 &   $<$0.01 & $\mp$0.06 & $+$0.01 & 0.11\\
Y II  &     $<$0.01 & $\pm$0.04 & $\pm$0.03 & $\mp$0.03 & $+$0.10 & 0.06\\
Zr II &     $<$0.01 & $\pm$0.04 & $\pm$0.03 & $\mp$0.01 & $+$0.09 & 0.06\\
Ba II &   $\pm$0.03 & $\pm$0.02 & $\pm$0.04 & $\mp$0.07 & $+$0.14 & 0.09\\
La II &   $\pm$0.02 & $\pm$0.04 & $\pm$0.03 &   $<$0.01 & $+$0.10 & 0.06\\
Ce II &   $\pm$0.01 & $\pm$0.04 & $\pm$0.03 & $\mp$0.01 & $+$0.09 & 0.06\\
Nd II &   $\pm$0.01 & $\pm$0.04 & $\pm$0.03 & $\mp$0.04 & $+$0.10 & 0.07\\
Eu II &     $<$0.01 & $\pm$0.04 & $\pm$0.03 & $\mp$0.01 & $+$0.10 & 0.06\\
\hline
\end{tabular}
\end{table*}
\clearpage
\begin{table*}[h!]
\renewcommand\thetable{A.2}
\caption{
LTE abundance results. Abundance ratios for ionized species are given relative to Fe\,{\sc ii}. 
For iron itself, [Fe/H] is listed. The line-to-line scatter, $\sigma$, and number of measured lines, $N$, indicate the statistical 
error.}
\centering
\begin{tabular}{cccccccccccccccc}
\hline\hline       
& \multicolumn{3}{c}{Hip 11575} &&  \multicolumn{3}{c}{Hip 9353} && \multicolumn{3}{c}{Hip 22336} && \multicolumn{3}{c}{Hip 22940} \\ 
\cline{2-4}\cline{6-8}\cline{10-12}\cline{14-16}
\rb{Species} & [X/Fe] & $\sigma$ & $N$ &&   [X/Fe] & $\sigma$ & $N$ &&  [X/Fe] & $\sigma$ & $N$  &&  [X/Fe] & $\sigma$ & $N$ \\
\hline       
C\,{\sc i}  & $-$0.04 &  0.12 &   2 && $-$0.20 &  0.03 &   2 && $-$0.19 &  0.06  &   2 &&   \phantom{$-$}0.02 &  0.09  &   2 \\ 
O\,{\sc i}  &  \phantom{$-$}0.10 &  0.02 &	3 &&   \phantom{$-$}0.06 &  0.05 &   3 &&   \phantom{$-$}0.02 &  0.02  &   3 &&   \phantom{$-$}0.06 &  0.03  &   3 \\ 
Na\,{\sc i}  & \phantom{$-$}0.19 &  0.06 &	4 &&   \phantom{$-$}0.25 &  0.12 &   4 &&   \phantom{$-$}0.02 &  0.10  &   4 &&   \phantom{$-$}0.14 &  0.17  &   4 \\ 
Mg\,{\sc i}  & \phantom{$-$}0.04 &  0.14 &	4 &&   \phantom{$-$}0.01 &  0.06 &   3 && $-$0.03 &  0.03  &   4 && $-$0.05 &  0.12  &   4 \\ 
Al\,{\sc i}  & \phantom{$-$}0.03 &  0.03 &	3 &&   \phantom{$-$}0.04 &  0.08 &   3 && $-$0.05 &  0.05  &   3 &&   \phantom{$-$}0.12 &  0.02  &   3 \\ 
Si\,{\sc i}  & $-$0.02 &  0.13 &  13 && $-$0.00 &  0.15 &  13 && $-$0.04 &  0.11  &  13 &&   \phantom{$-$}0.01 &  0.10  &  13 \\ 
S\,{\sc i}  & $-$0.06 &  0.07 &   5 && $-$0.15 &  0.07 &   5 && $-$0.20 &  0.14  &   5 && $-$0.08 &  0.06  &   5 \\ 
K\,{\sc i}  &  \phantom{$-$}0.44 &  \dots &   1 &&   \phantom{$-$}0.46 &  \dots &   1 &&   \phantom{$-$}0.33 &  \dots  &   1 &&   \phantom{$-$}0.13 &  \dots  &   1 \\ 
Ca\,{\sc i}  & $-$0.01 &  0.09 &  15 &&   \phantom{$-$}0.02 &  0.16 &  15 &&   \phantom{$-$}0.00 &  0.13  &  15 &&   \phantom{$-$}0.01 &  0.12  &  14 \\ 
Sc\,{\sc i}  & $-$0.05 &  0.10 &   4 &&   \phantom{$-$}0.06 &  0.18 &   5 &&   \phantom{$-$}0.07 &  0.16  &   5 &&   \phantom{$-$}0.19 &  0.10  &   6 \\ 
Ti\,{\sc i}  & $-$0.08 &  0.11 &  23 &&   \phantom{$-$}0.02 &  0.16 &  22 && \phantom{$-$}0.00 &  0.12  &  23 &&   \phantom{$-$}0.15 &  0.16  &  23 \\ 
Ti\,{\sc ii} & \phantom{$-$}0.09 &  0.10 &	4 &&   \phantom{$-$}0.09 &  0.19 &   6 &&   \phantom{$-$}0.23 &  0.17  &   4 &&  \phantom{$-$}0.26 &  0.21  &   5 \\ 
V\,{\sc i}  & $-$0.06 &  0.02 &   4 &&   \phantom{$-$}0.01 &  0.09 &   4 && $-$0.10 &  0.06  &   4 &&  \phantom{$-$}0.14 &  0.20  &   4 \\ 
Cr\,{\sc i}  & $-$0.05 &  0.23 &   9 &&  \phantom{$-$}0.07 &  0.13 &   7 && $-$0.01 &  0.17  &   8 &&   \phantom{$-$}0.02 &  0.06  &   8 \\ 
Mn\,{\sc i}  & $-$0.08 &  0.09 &   6 && $-$0.06 &  0.07 &   6 && $-$0.17 &  0.10  &   6 &&  \phantom{$-$}0.02 &  0.17  &   4 \\ 
Fe\,{\sc i}  & \phantom{$-$}0.20 &  0.12 & 111 &&   \phantom{$-$}0.32 &  0.10 & 112 &&  \phantom{$-$}0.30 &  0.12  & 113 &&   \phantom{$-$}0.17 &  0.13  & 102 \\ 
Fe\,{\sc ii} & \phantom{$-$}0.20 &  0.11 &  16 &&   \phantom{$-$}0.31 &  0.12 &  16 &&  \phantom{$-$}0.30 &  0.11  &  15 &&   \phantom{$-$}0.17 &  0.10  &  14 \\ 
Co\,{\sc i}  & \phantom{$-$}0.06 &  0.14 &  11 &&   \phantom{$-$}0.06 &  0.19 &  11 && $-$0.07 &  0.12  &  11 &&   \phantom{$-$}0.14 &  0.31  &  11 \\ 
Ni\,{\sc i}  & \phantom{$-$}0.02 &  0.18 &  26 && $-$0.01 &  0.12 &  26 && $-$0.03 &  0.11  &  26 &&   \phantom{$-$}0.08 &  0.16  &  25 \\ 
Cu\,{\sc i}  & \phantom{$-$}0.05 &  0.10 &	3 &&   \phantom{$-$}0.17 &  0.10 &   3 && $-$0.11 &  0.09  &   3 &&  \phantom{$-$}0.07 &  0.23  &   3 \\ 
Y\,{\sc ii} & $-$0.12 &  0.23 &   3 && $-$0.16 &  0.10 &   3 && $-$0.18 &  0.02  &   3 && $-$0.04 &  0.17  &   3 \\ 
Zr\,{\sc ii} & $-$0.12 &  \dots &   1 && $-$0.35 &  \dots &   1 && $-$0.31 &  \dots  &   1 && $-$0.14 &  \dots  &   1 \\ 
Ba\,{\sc ii} & $-$0.19 &  0.04 &	3 &&   $-$0.07 &  0.08 &   3 &&   $-$0.03 &  0.06  &   3 &&   \phantom{$-$}0.08 &  0.13  &   3 \\ 
La\,{\sc ii} & $-$0.02 &  0.01 &   2 &&  \phantom{$-$}0.13 &  0.04 &   2 &&  \phantom{$-$}0.20 &  0.01  &   2 &&  \phantom{$-$}0.37 &  0.18  &   2 \\ 
Ce\,{\sc ii} & $-$0.01 &  \dots &   1 && $-$0.20 &  \dots &   1 && \phantom{$-$}0.02 &  \dots  &   1 &&   \phantom{$-$}0.24 &  \dots  &   1 \\ 
Eu\,{\sc ii} & \phantom{$-$}0.10 &  \dots &   1 && $-$0.10 &  \dots &   1 && $-$0.32 &  \dots  &   1 &&   \phantom{$-$}0.01 &  \dots  &   1 \\ %
$[hs/ls]$\tablefootmark{a} & \phantom{$-$}0.27 & \dots & \dots && \phantom{$-$}0.82 & \dots & \dots && \phantom{$-$}0.99 & \dots & \dots && \phantom{$-$}0.86 & \dots & \dots \\
\hline                  
& \multicolumn{3}{c}{Hip 23311} && \multicolumn{3}{c}{Hip 26834} && \multicolumn{3}{c}{Intrinscic Scatter} &&&& \\ 
\cline{2-4}\cline{6-8}\cline{10-12}
\rb{Species} & [X/Fe] & $\sigma$ & $N$ &&  [X/Fe] & $\sigma$ & $N$ && \multicolumn{3}{c}{$\sigma_0$} &&&&\\
\hline                  
C\,{\sc i}   & \phantom{$-$}0.28 &  0.09 &   2 && $-$0.18 &  0.05 &   2 &&    & 0.25$\pm$0.07 &&&&& \\ 
O\,{\sc i}   & \phantom{$-$}0.20 &  0.16 &   3 && $-$0.08 &  0.05 &   3 &&    & 0.06$\pm$0.02 &&&&& \\ 
Na\,{\sc i}   & \phantom{$-$}0.39 &  0.23 &   4 &&   \phantom{$-$}0.15 &  0.13 &   4 &&	   & 0.07$\pm$0.04 &&&&& \\ 
Mg\,{\sc i}   & \phantom{$-$}0.18 &  0.06 &   4 &&   \phantom{$-$}0.24 &  0.36 &   5 &&	   & 0.08$\pm$0.03 &&&&& \\ 
Al\,{\sc i}   & \phantom{$-$}0.19 &  0.16 &   3 &&  \phantom{$-$}0.11 &  0.05 &   3 &&	   & 0.07$\pm$0.03 &&&&& \\ 
Si\,{\sc i}   & \phantom{$-$}0.15 &  0.14 &  13 && $-$0.03 &  0.12 &  13 &&    & 0.05$\pm$0.02 &&&&& \\ 
 S\,{\sc i}   & $-$0.24 &  0.25 &   3 && $-$0.13 &  0.04 &   4 &&  & 0.02$\pm$0.02 &&&&& \\ 
 K\,{\sc i}   & \phantom{$-$}0.17 &  \dots &   1 &&   \phantom{$-$}0.16 &  \dots &   1 &&    & 0.14$\pm$0.04 &&&&& \\ 
Ca\,{\sc i}   & \phantom{$-$}0.09 &  0.06 &   4 && $-$0.01 &  0.14 &  13 &&    & 0.02$\pm$0.02 &&&&& \\ 
Sc\,{\sc i}   & \phantom{$-$}0.10 &  0.20 &   5 &&   \phantom{$-$}0.10 &  0.12 &   5 &&	   & 0.06$\pm$0.03 &&&&& \\ 
Ti\,{\sc i}   & \phantom{$-$}0.30 &  0.13 &  22 &&   \phantom{$-$}0.04 &  0.11 &  22 &&	   & 0.12$\pm$0.04 &&&&& \\ 
Ti\,{\sc ii}  & \phantom{$-$}0.11 &  0.13 &   6 &&   \phantom{$-$}0.27 &  0.17 &   4 &&	   & 0.32$\pm$0.16 &&&&& \\ 
V\,{\sc i}   &  \phantom{$-$}0.32 &  0.17 &   4 &&   \phantom{$-$}0.02 &  0.15 &   4 &&	   & 0.12$\pm$0.05 &&&&& \\ 
Cr\,{\sc i}   & \phantom{$-$}0.07  &  0.07 &   7 &&  \phantom{$-$}0.00  &  0.15 & 8  &&      & 0.17$\pm$0.09 &&&&& \\ 
Mn\,{\sc i}   & \phantom{$-$}0.10 &  0.17 &   6 && $-$0.03 &  0.15 &   6 &&    & 0.05$\pm$0.03 &&&&& \\ 
Fe\,{\sc i}   & \phantom{$-$}0.13 &  0.17 & 115 &&   \phantom{$-$}0.38 &  0.13 & 108 &&	   & 0.09$\pm$0.03 &&&&& \\ 
Fe\,{\sc ii}  & \phantom{$-$}0.14 &  0.15 &  13 &&   \phantom{$-$}0.38 &  0.16 &  16 &&	   & 0.08$\pm$0.03 &&&&& \\ 
Co\,{\sc i}   & \phantom{$-$}0.18 &  0.17 &  10 &&   \phantom{$-$}0.01 &  0.16 &   9 &&	   & 0.07$\pm$0.03 &&&&& \\ 
Ni\,{\sc i}   & \phantom{$-$}0.06 &  0.20 &  25 &&   \phantom{$-$}0.06 &  0.15 &  27 &&	   & 0.03$\pm$0.02 &&&&& \\ 
Cu\,{\sc i}   & \phantom{$-$}0.11 &  0.23 &   2 &&   \phantom{$-$}0.04 &  0.18 &   3 &&	   & 0.08$\pm$0.04 &&&&& \\ 
Y\,{\sc ii}  & \phantom{$-$} 0.41 &  \dots &   1 && $-$0.09 &  0.16 &   3 &&   & 0.21$\pm$0.06 &&&&& \\ 
Zr\,{\sc ii}  & \phantom{$-$}0.39 &  \dots &   1 && $-$0.31 &  \dots &   1 &&  & 0.25$\pm$0.07 &&&&& \\ 
Ba\,{\sc ii}  & \phantom{$-$}0.18 &  0.19 &   3 &&   \phantom{$-$}0.04 &  0.09 &   3 &&	   & 0.08$\pm$0.03 &&&&& \\ 
La\,{\sc ii}  & \phantom{$-$}0.25 &  \dots &   1 &&   \phantom{$-$}0.32 &  0.07 &   2 &&     & 0.12$\pm$0.04 &&&&& \\ 
Ce\,{\sc ii}  & \phantom{$-$}0.24 &  \dots &   1 &&   \phantom{$-$}0.01 &  \dots &   1 &&    & 0.16$\pm$0.04 &&&&& \\ 
Eu\,{\sc ii}  & $-$0.01 &  \dots &   1 &&  \phantom{$-$}0.05 &  \dots &   1 &&  & 0.14$\pm$0.04 &&&&& \\ 
$[hs/ls]$\tablefootmark{a} & $-$0.40 & \dots & \dots && \phantom{$-$}0.97 & \dots & \dots && & & &&  & & \\\hline
\end{tabular}
\tablefoot{\tablefoottext{a}{$[hs/ls]=[<$Ba,La$>$/$<$Y,Zr$>]$}}
\end{table*}
\clearpage
\begin{table*}[h!] \label{tab:orbital_parameters}
  \renewcommand\thetable{B.1}
\caption{Observed and derived kinematic properties} 
\renewcommand{\arraystretch}{1.5}
\centering
\begin{tabular}{rcccccc}
\hline\hline       
Parameter  & Hip\,11575 & Hip\,9353 & Hip\,22336 & Hip\,22940 & Hip\,23311 & Hip\,26834  \\
\hline       
Distance [pc] & \phantom{$-$}$98.18 \pm 0.14$ & \phantom{$-$2}$31.75 \pm 0.06$ & \phantom{$-$2}$26.12 \pm 0.02$ & \phantom{$-$2}$36.03 \pm 0.03$ & \phantom{$-$110}$8.84 \pm 0.002$ & \phantom{$-$1}$30.06 \pm 0.02$ \\
$\mu_\alpha$ $[\text{mas yr}^{-1}]$ & \phantom{$-$}$99.81 \pm 0.01$ & \phantom{$-$}$230.83 \pm 0.14$ & \phantom{$-$}$311.41 \pm 0.04$ & \phantom{$-$}$151.02 \pm 0.03$ & \phantom{$-$1}$549.31 \pm 0.02$ & \phantom{$-$}$243.76 \pm 0.02$ \\
$\mu_\delta$ $[\text{mas yr}^{-1}]$ & $-53.79 \pm 0.06$ & $-256.47 \pm 0.05$ & $-248.83 \pm 0.04$ & $-292.11 \pm 0.02$ & $-1108.25 \pm 0.02$ & $-117.81 \pm 0.02$ \\
v$_{\rm HC}$ [km s$^{-1}$] & \phantom{$-$}$37.17 \pm 0.12$ & $-18.50 \pm 0.14$ & \phantom{$-$2}$77.12 \pm 0.12$ & \phantom{$-$2}$14.96 \pm 0.12$ & \phantom{$-$11}$21.36 \pm 0.12$ & \phantom{$-$1}$59.16 \pm 0.12$ \\

\hline

$\text{X}$ $[\text{pc}]$ &
\phantom{4}$-9.9$ & $-16.2$ & \phantom{4}$-20.8$ & \phantom{4}$-31.7$ & $-7.1$ & $-21.4$\\
$\text{Y}$ $[\text{pc}]$ &
$-40.1$ & \phantom{$-$1}$7.8$ & \phantom{10}$-9.0$ & \phantom{4}$-9.1$ & $-3.3$ & $-17.8$\\
$\text{Z}$ $[\text{pc}]$ &
$-89.1$ & $-26.2$ & \phantom{4}$-13.0$ & $-14.6$ & $-4.0$ & $-11.4$\\

$\text{U}_{\text{LSR}}$ $[\text{km s}^{-1}]$ &
\phantom{4}$-4.78\pm1.26$ &
\phantom{$-$}$12.86\pm1.25$ &
$-41.19\pm1.25$ &
\phantom{$-$}$15.05\pm1.25$ &
\phantom{$-$}$16.35\pm1.25$ &
$-19.72\pm1.25$\\

$\text{V}_{\text{LSR}}$ $[\text{km s}^{-1}]$ &
$-49.25\pm2.06$ &
$-42.64\pm2.06$ &
$-59.81\pm2.06$ &
$-44.93\pm2.06$ &
$-42.54\pm2.06$ &
$-51.26\pm2.06$\\

$\text{W}_{\text{LSR}}$ $[\text{km s}^{-1}]$ &
\phantom{4}$-4.24\pm0.64$ &
\phantom{$-$}$12.29\pm0.64$ &
$-14.19\pm0.63$ &
\phantom{4}$-2.52\pm0.63$ &
\phantom{$-$}$-3.50\pm0.63$ &
\phantom{$-$5}$8.18\pm0.62$\\

\hline

$\text{E}$ $[\text{km}^2 \, \text{s}^{-2}]$ &
$-183738^{+11}_{-11}$ &
$-184402^{+12}_{-12}$ &
$-181499^{+10}_{-10}$ &
$-184178^{+9}_{-9}$ &
$-184421^{+4}_{-4}$ &
$-183320^{+12}_{-12}$\\

$\text{E}_{\text{J}}$ $[\text{km}^2 \, \text{s}^{-2}]$ &
$-1.837^{+<10^{-3}}_{-<10^{-3}}$ &
$-1.844^{+<10^{-3}}_{-<10^{-3}}$ &
$-1.815^{+<10^{-3}}_{-<10^{-3}}$ &
$-1.842^{+<10^{-3}}_{-<10^{-3}}$ &
$-1.844^{+<10^{-3}}_{-<10^{-3}}$ &
$-1.833^{+<10^{-3}}_{-<10^{-3}}$\\

$\text{L}_{\text{Z}}$ $[\text{kpc km s}^{-1}]$ &
$1491.8^{+0.7}_{-0.7}$ &
$1546.9^{+0.9}_{-0.9}$ &
$1408.4^{+0.4}_{-0.4}$ &
$1530.9^{+0.7}_{-0.7}$ &
$1545.9^{+0.4}_{-0.4}$ &
$1478.4^{+0.7}_{-0.7}$\\

$\text{V}_{\text{R}}$ $[\text{km s}^{-1}]$ &
\phantom{$-$}$3.93^{+0.03}_{-0.03}$ &
$-12.69^{+0.10}_{-0.10}$ &
\phantom{$-$}$41.09^{+0.09}_{-0.09}$ &
$-15.21^{+0.12}_{-0.12}$ &
$-16.32^{+0.11}_{-0.11}$ &
\phantom{$-$}$19.28^{+0.11}_{-0.11}$\\

$\text{V}_{\text{T}}$ $[\text{km s}^{-1}]$ &
\phantom{$-$}$183.44^{+0.09}_{-0.09}$ &
\phantom{$-$}$190.07^{+0.11}_{-0.11}$ &
\phantom{$-$}$172.96^{+0.04}_{-0.04}$ &
\phantom{$-$}$187.75^{+0.09}_{-0.09}$ &
\phantom{$-$}$190.17^{+0.05}_{-0.05}$ &
\phantom{$-$}$181.55^{+0.08}_{-0.08}$\\

$\text{R}_{\text{Apo}}$ $[\text{kpc}]$ &
$8.43^{+<0.01}_{-<0.01}$ &
$8.54^{+<0.01}_{-<0.01}$ &
$8.57^{+<0.01}_{-<0.01}$ &
$8.60^{+<0.01}_{-<0.01}$ &
$8.58^{+<0.01}_{-<0.01}$ &
$8.46^{+<0.01}_{-<0.01}$\\

$\text{R}_{\text{Peri}}$ $[\text{kpc}]$ &
$3.72^{+<0.01}_{-<0.01}$ &
$3.72^{+<0.01}_{-<0.01}$ &
$3.37^{+<0.01}_{-<0.01}$ &
$3.67^{+<0.01}_{-<0.01}$ &
$3.69^{+<0.01}_{-<0.01}$ &
$3.66^{+<0.01}_{-<0.01}$\\

$e$ &
$0.39^{+<0.01}_{-<0.01}$ &
$0.39^{+<0.01}_{-<0.01}$ &
$0.44^{+<0.01}_{-<0.01}$ &
$0.40^{+<0.01}_{-<0.01}$ &
$0.40^{+<0.01}_{-<0.01}$ &
$0.40^{+<0.01}_{-<0.01}$\\

$\text{Z}_{\text{max}}$ $[\text{pc}]$ &
$91.6^{+0.9}_{-0.9}$ &
$186.1^{+5.8}_{-5.8}$ &
$237.9^{+0.5}_{-0.5}$ &
$36.6^{+0.8}_{-0.8}$ &
$56.0^{+1.1}_{-1.1}$ &
$126.1^{+0.4}_{-0.4}$\\

$\text{P}_{\text{z}}$ $[\text{Gyr}]$ &
$0.063^{+<10^{-3}}_{-<10^{-3}}$ &
$0.066^{+<10^{-3}}_{-<10^{-3}}$ &
$0.067^{+<10^{-3}}_{-<10^{-3}}$ &
$0.061^{+<10^{-3}}_{-<10^{-3}}$ &
$0.061^{+<10^{-3}}_{-<10^{-3}}$ &
$0.065^{+<10^{-3}}_{-<10^{-3}}$\\

\hline                  

$\Omega_{R,\mathrm{ini}}$ [Gyr$^{-1}$] &
$49.08^{+<0.01}_{-<0.01}$ &
$48.84^{+<0.01}_{-<0.01}$ &
$56.24^{+<0.01}_{-<0.01}$ &
$48.85^{+<0.01}_{-<0.01}$ &
$48.83^{+<0.01}_{-<0.01}$ &
$49.13^{+<0.01}_{-<0.01}$\\

$\Omega_{\phi,\mathrm{ini}}$ [Gyr$^{-1}$] &
$38.35^{+<0.01}_{-<0.01}$ &
$38.34^{+<0.01}_{-<0.01}$ &
$38.35^{+<0.01}_{-<0.01}$ &
$38.34^{+<0.01}_{-<0.01}$ &
$38.34^{+<0.01}_{-<0.01}$ &
$38.35^{+<0.01}_{-<0.01}$\\

$\Omega_{z,\mathrm{ini}}$ [Gyr$^{-1}$] &
$92.54^{+0.05}_{-0.05}$ &
$107.01^{+0.10}_{-0.10}$ &
$98.17^{+0.09}_{-0.09}$ &
$115.34^{+0.03}_{-0.03}$ &
$114.48^{+0.03}_{-0.03}$ &
$104.02^{+0.02}_{-0.02}$\\

$\Omega_{R,\mathrm{bar}}$ [Gyr$^{-1}$] &
$49.08^{+<0.01}_{-<0.01}$ &
$48.84^{+<0.01}_{-<0.01}$ &
$56.24^{+<0.01}_{-<0.01}$ &
$48.85^{+<0.01}_{-<0.01}$ &
$48.83^{+<0.01}_{-<0.01}$ &
$49.13^{+<0.01}_{-<0.01}$\\

$\Omega_{\phi,\mathrm{bar}}$ [Gyr$^{-1}$] &
$-38.36^{+<0.01}_{-<0.01}$ &
$-38.34^{+<0.01}_{-<0.01}$ &
$-38.35^{+<0.01}_{-<0.01}$ &
$-38.34^{+<0.01}_{-<0.01}$ &
$-38.34^{+<0.01}_{-<0.01}$ &
$-38.35^{+<0.01}_{-<0.01}$\\

$\Omega_{z,\mathrm{bar}}$ [Gyr$^{-1}$] &
$92.54^{+0.05}_{-0.05}$ &
$107.01^{+0.10}_{-0.10}$ &
$98.17^{+0.09}_{-0.09}$ &
$115.34^{+0.03}_{-0.03}$ &
$114.48^{+0.03}_{-0.03}$ &
$104.02^{+0.02}_{-0.02}$\\

$\Delta\Omega_{R}$ &
$0.00^{+<0.01}_{-<0.01}$ &
$0.00^{+<0.01}_{-<0.01}$ &
$0.00^{+<0.01}_{-<0.01}$ &
$0.00^{+<0.01}_{-<0.01}$ &
$0.00^{+<0.01}_{-<0.01}$ &
$0.00^{+0.12}_{-0.12}$\\

$\Delta\Omega_{\phi}$ &
$0.00^{+<0.01}_{-<0.01}$ &
$0.00^{+<0.01}_{-<0.01}$ &
$0.00^{+<0.01}_{-<0.01}$ &
$0.00^{+<0.01}_{-<0.01}$ &
$0.00^{+<0.01}_{-<0.01}$ &
$0.00^{+<0.01}_{-<0.01}$\\

$\Delta\Omega_{z}$ &
$0.13^{+<0.01}_{-<0.01}$ &
$0.19^{+<0.01}_{-<0.01}$ &
$0.01^{+<0.01}_{-<0.01}$ &
$0.00^{+<0.01}_{-<0.01}$ &
$0.00^{+<0.01}_{-<0.01}$ &
$0.15^{+<0.01}_{-<0.01}$\\

$\log{\Delta\Omega}$ &
$-0.89^{+<0.01}_{-<0.01}$ &
$-0.72^{+0.01}_{-0.01}$ &
$-2.21^{+0.04}_{-0.04}$ &
$-2.98^{+<0.01}_{-<0.01}$ &
$-2.96^{+0.03}_{-0.03}$ &
$-0.82^{+0.11}_{-0.11}$\\
\hline   

\end{tabular}
\end{table*}
\clearpage
\begin{figure*}
  \renewcommand\thefigure{B.1}
    \centering
    \includegraphics[width=0.47\linewidth]{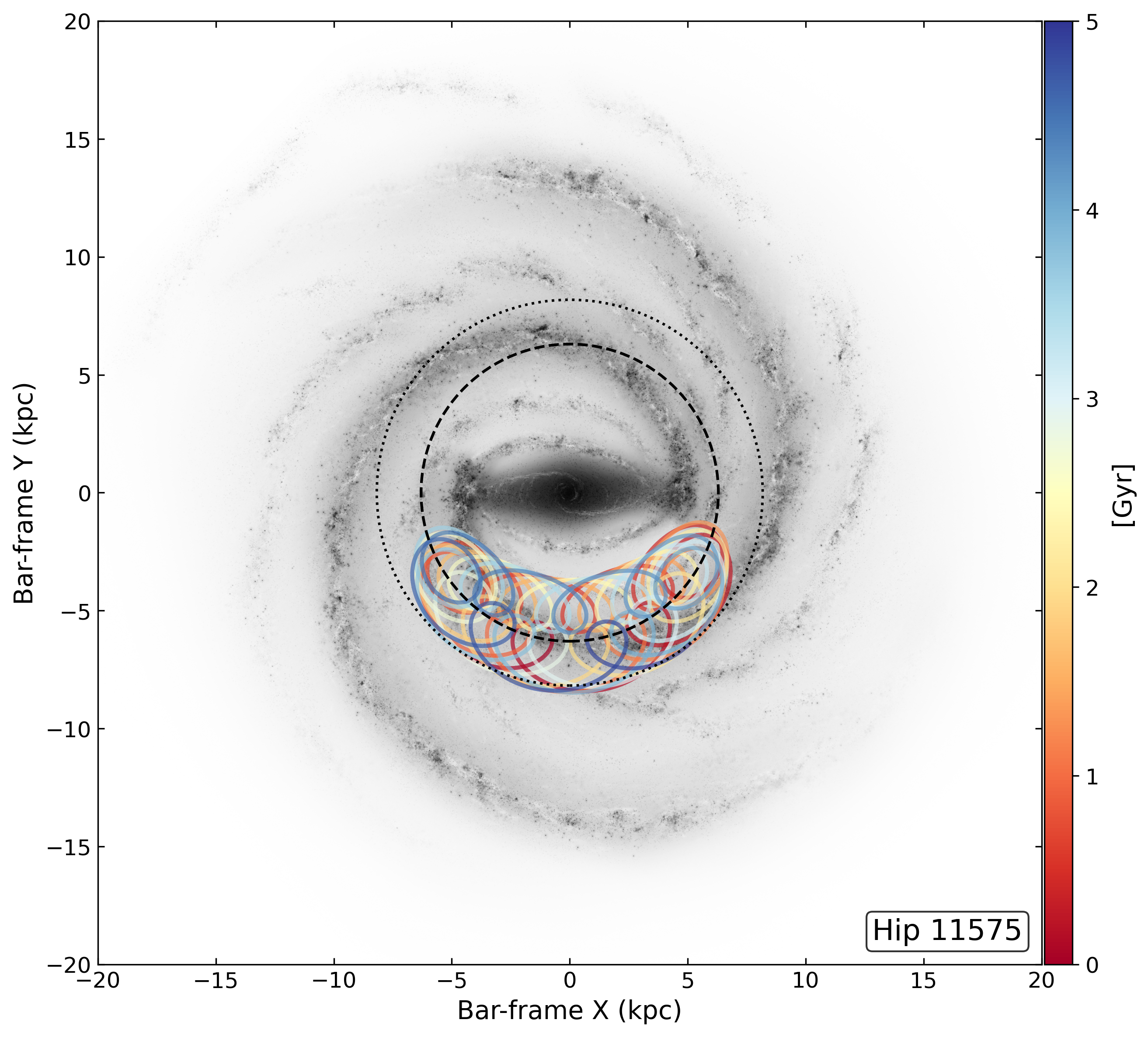}
    \includegraphics[width=0.47\linewidth]{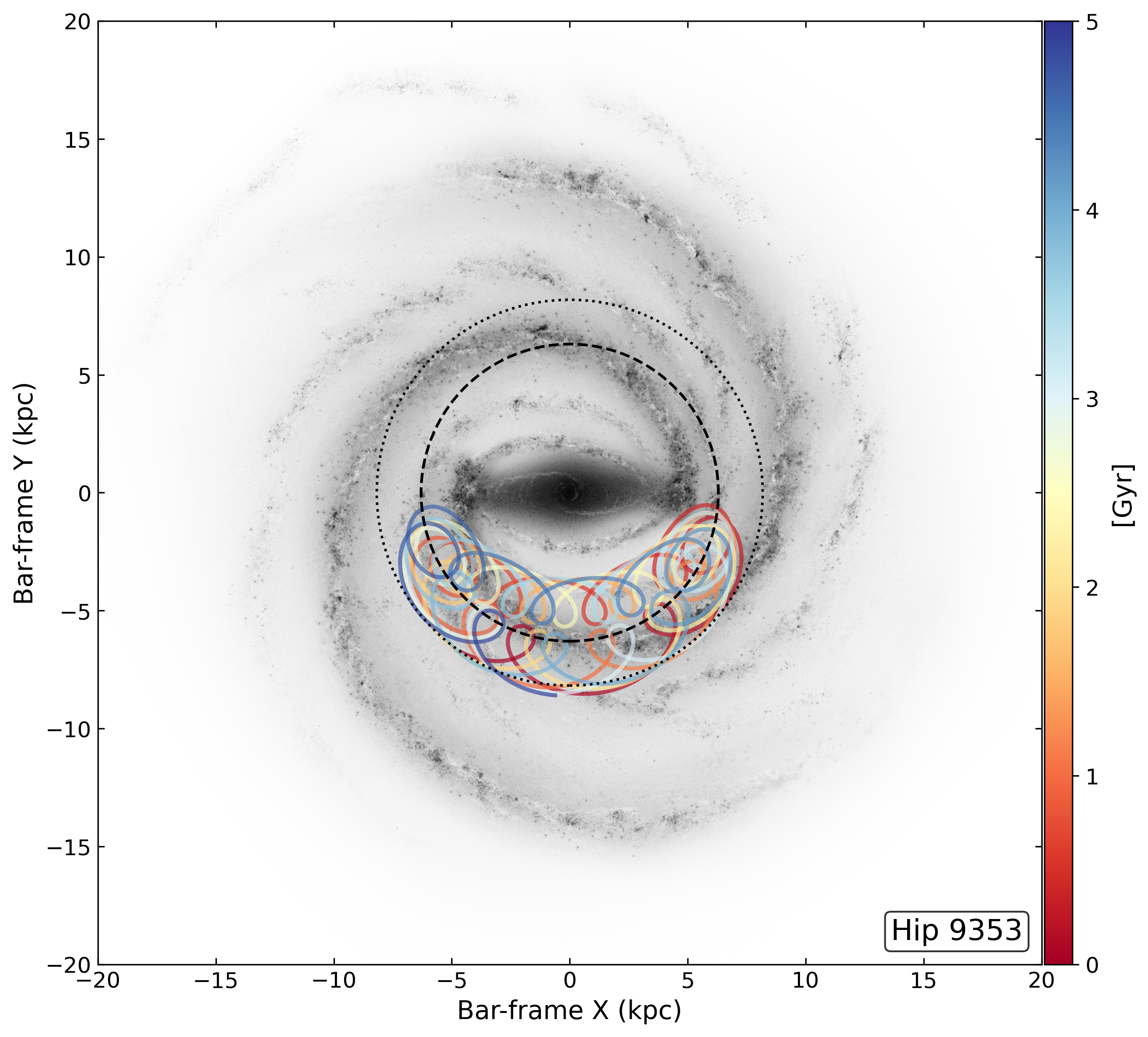}
    \includegraphics[width=0.47\linewidth]{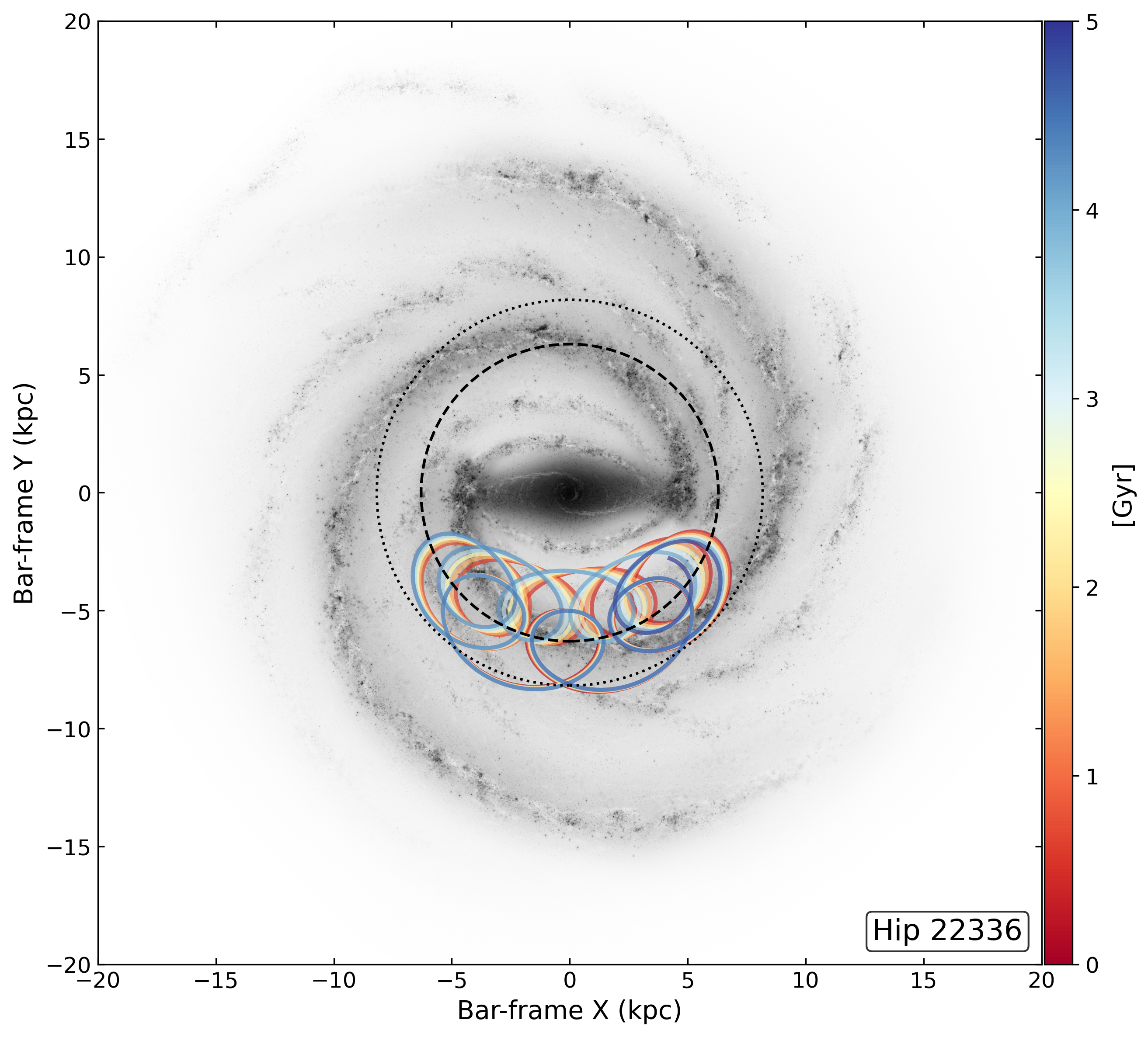}
    \includegraphics[width=0.47\linewidth]{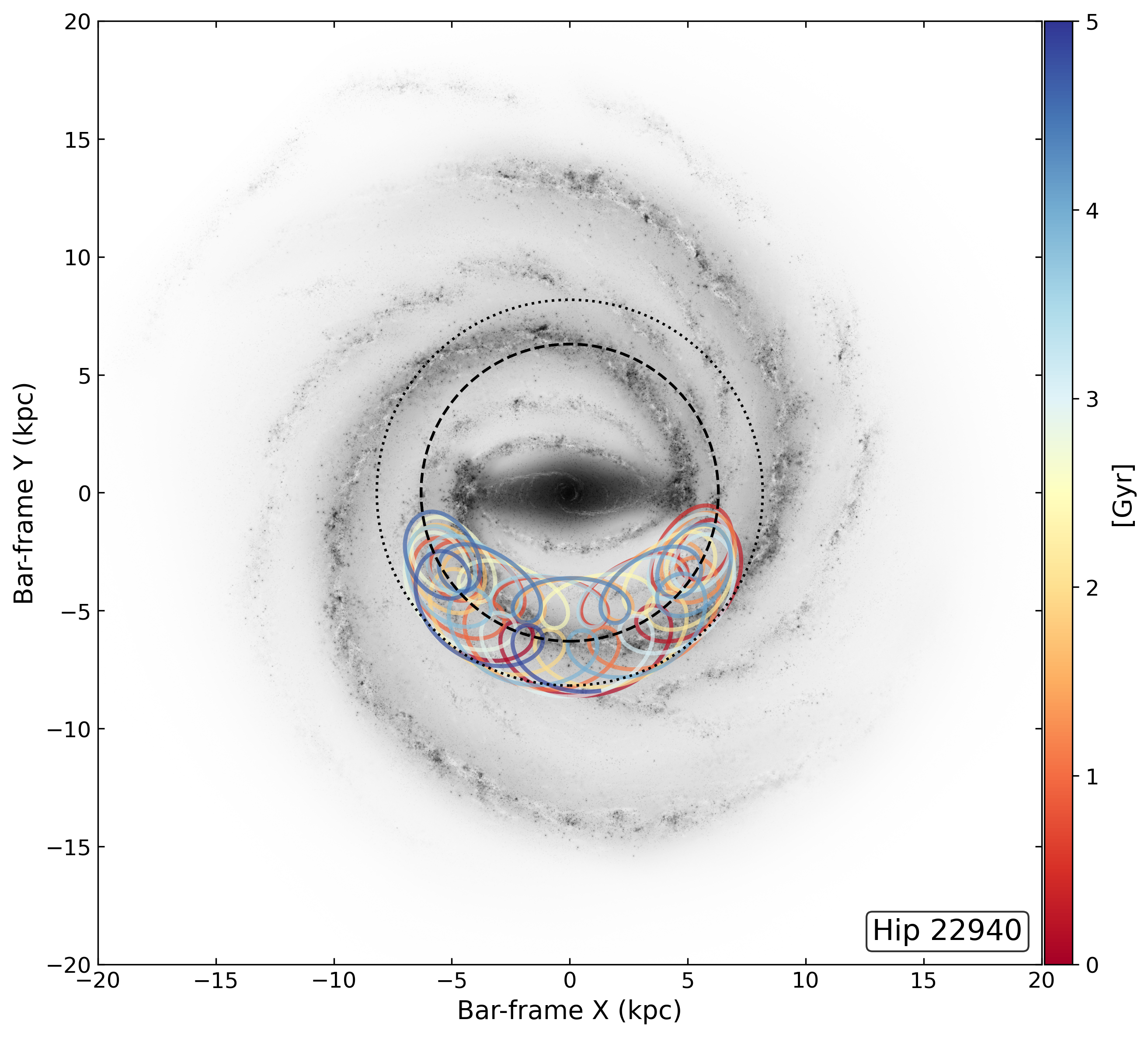}
    \includegraphics[width=0.47\linewidth]{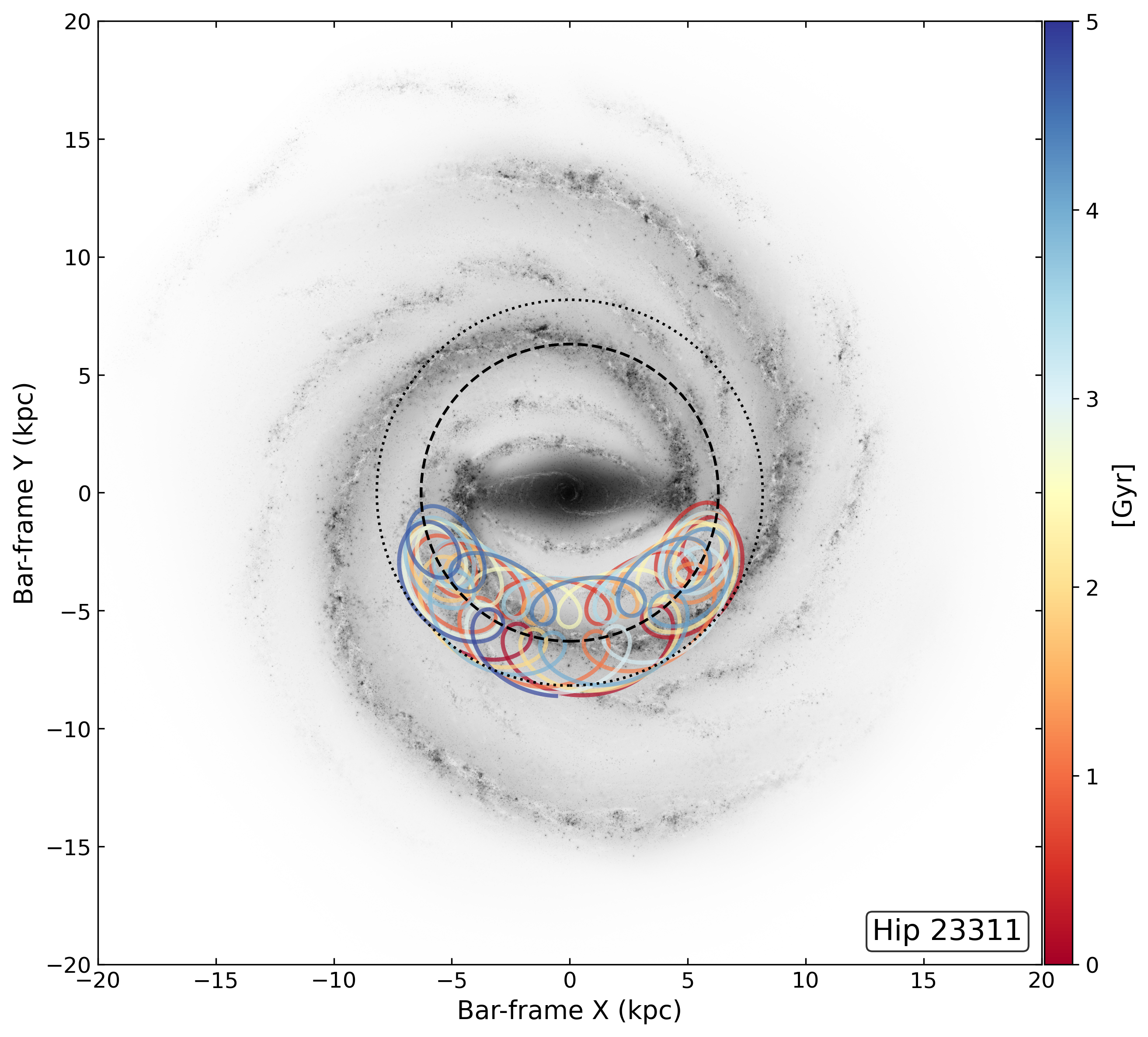}
    \includegraphics[width=0.47\linewidth]{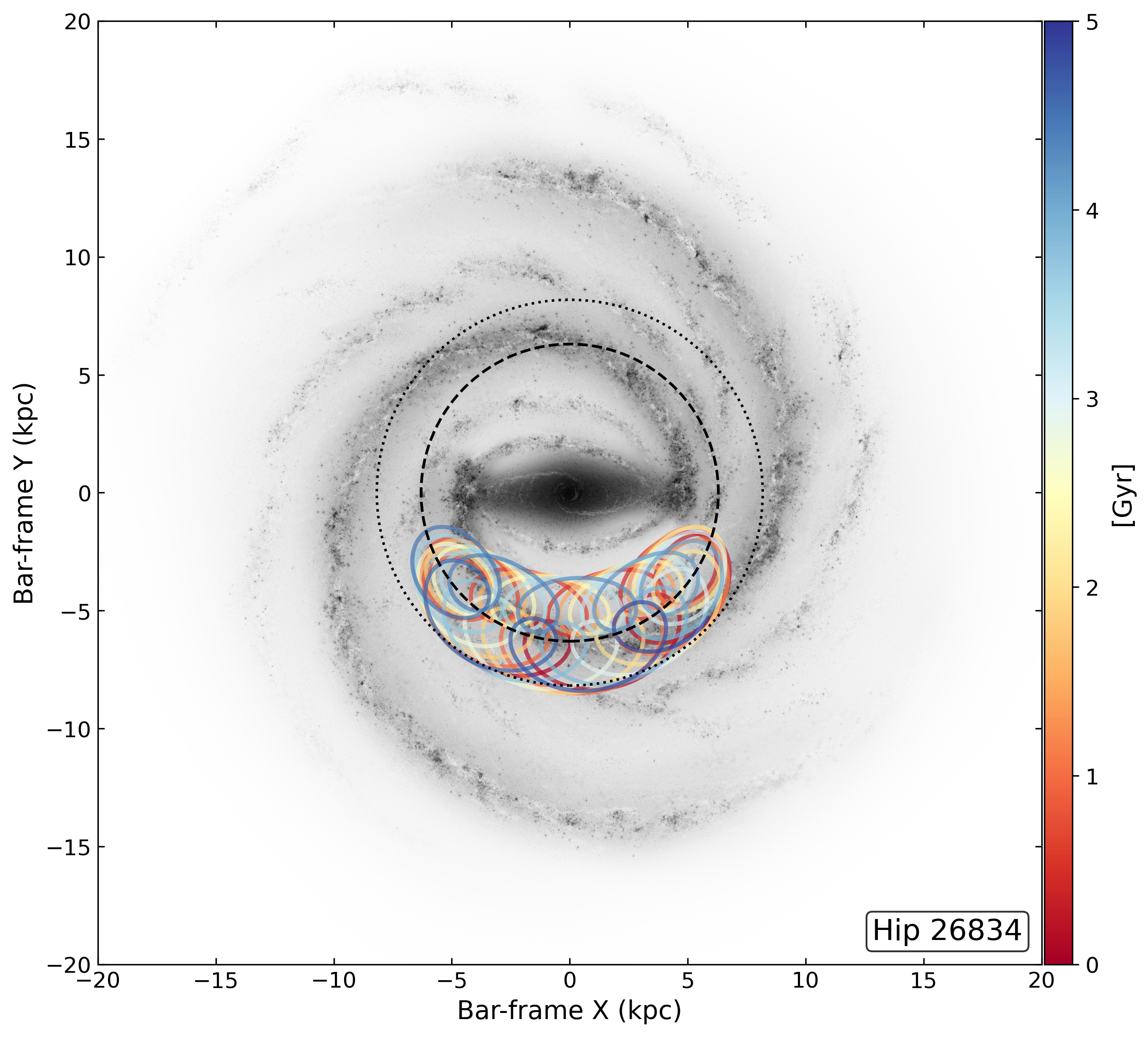}
    \caption{Face-on view of the Galactic disk showing the orbits of the six HR~1614 MG stars in the rotating bar frame.
    The grayscale background shows a schematic MW disk, including the bar and spiral arms, from the \texttt{mw-plot} package \citep{mw_plot}. Colored curves trace the stellar orbits over the past $5,\text{Gyr}$ in the
    potential with a rigidly rotating bar of pattern speed $\Omega_{\rm bar}=37,\text{km s}^{-1},\text{kpc}^{-1}$. The color encodes the time along each orbit from 0 to $5,\text{Gyr}$  and the dashed line
    indicates the CR radius at 6.3 kpc, while the dotted circle marks the solar radius at $R_\odot=8.217\,\text{kpc}$.}

\end{figure*}
\clearpage
\section{Time evolution of Galactocentric radii}
\label{app:rgc_time_evolution}

\begin{figure}
  \renewcommand\thefigure{B.2}
    \centering
    \includegraphics[width=1\linewidth]{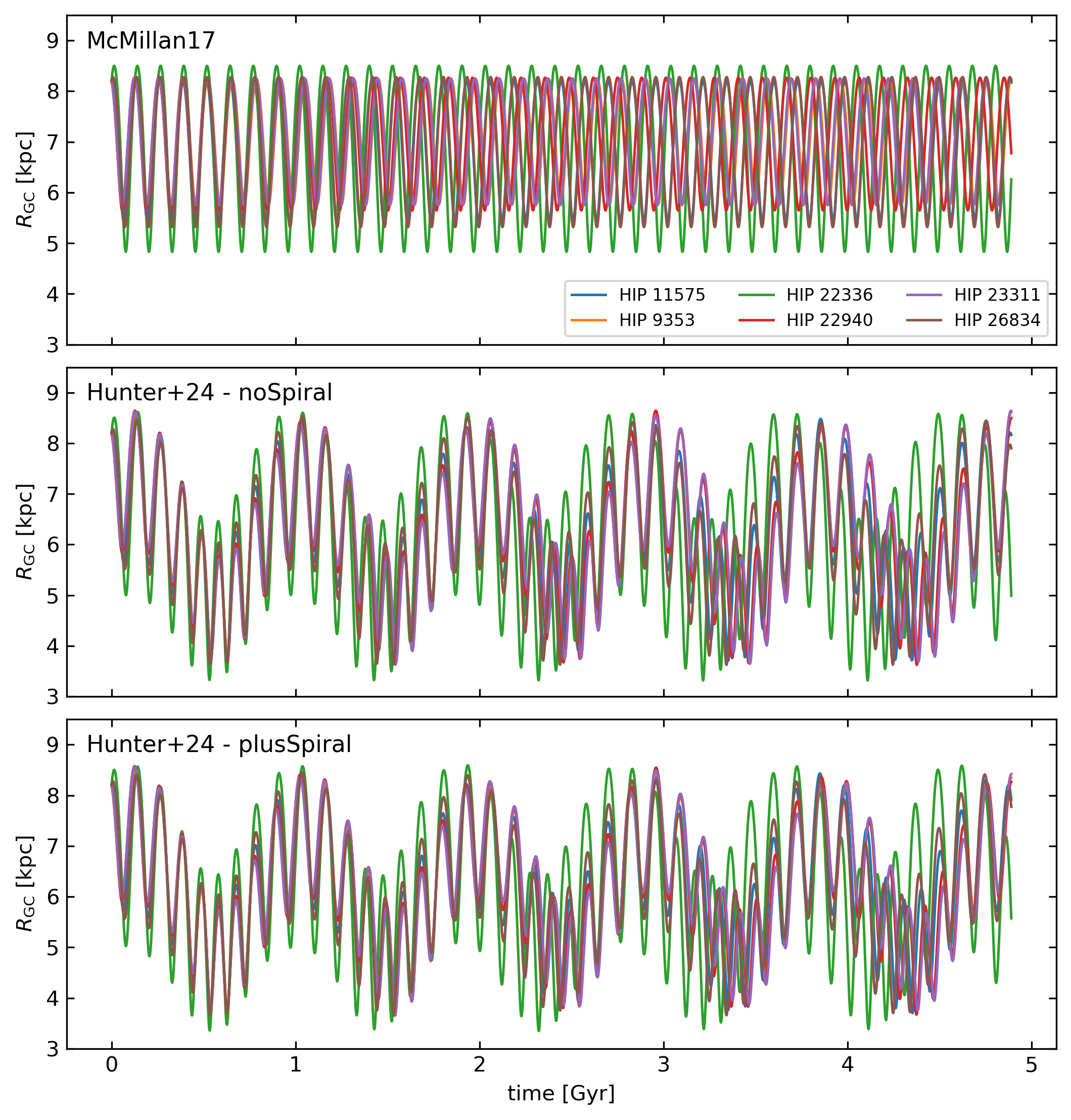}
    \caption{ 
    Time evolution of the Galactocentric radius, $R_{\rm GC}$, for the six stars in the axisymmetric potential
    (top panel), the barred potential without spiral arms 
    (middle panel), and the barred potential including spiral arms (bottom panel).
    }
    \label{fig:rgc_time_potential_comparison}
\end{figure}

{
To quantify the radial extent of the orbits and their dependence on details in the potentials, we illustrate here 
the time evolution of the Galactocentric radius,  
$R_{\rm GC}$, for the six stars in the three different  considered Galactic potentials: the axisymmetric potential of \citet{McMillan2017}, the nonaxisymmetric potential of \citet{Hunter2024} (labeled ``H24'' in Table~B.2)  without spiral arms, and the latter, but including spiral arms.
The resulting 
curves are shown in Fig.~\ref{fig:rgc_time_potential_comparison}, while the corresponding mean, median, minimum, and maximum radii are listed in Table~\ref{tab:rgc_statistics}.

In the axisymmetric potential \citep{McMillan2017}, the stars remain on 
regular radial oscillations around the solar radius.
The mean Galactocentric radii lie in a narrow range from $6.88$ to $7.10\,{\rm kpc}$, with an average over the six stars of $\langle R_{\rm GC} \rangle = 7.00\,{\rm kpc}$.
The corresponding median radii are slightly larger, with an average value of $7.13\,{\rm kpc}$.
The radial excursions are moderate, reaching minimum radii between $4.83$ and $5.77\,{\rm kpc}$ and maximum radii between $8.19$ and $8.50\,{\rm kpc}$.

The inclusion of the Galactic bar substantially changes the radial behaviour.
In the  potential of \citet{Hunter2024} without spiral arms, the mean radii decrease to $\langle R_{\rm GC} \rangle = 6.20\,{\rm kpc}$, while the median radii decrease to $6.14\,{\rm kpc}$.
This corresponds to a reduction of about $0.8\,{\rm kpc}$ in the mean radius and about $1.0\,{\rm kpc}$ in the median radius compared to the axisymmetric case.
At the same time, the minimum radii become significantly smaller, reaching $R_{\rm GC,min} \simeq 3.3$--$3.7\,{\rm kpc}$.
The maximum radii remain close to, or slightly larger than, those in the axisymmetric model, with values around $8.5$--$8.6\,{\rm kpc}$.
Thus, the lower mean and median radii in the barred model are primarily driven by repeated inward radial excursions rather than by a uniform shift of the entire orbit to smaller radii.

Adding spiral arms to the barred potential has only a minor effect on these results.
The mean and median radii in the barred model with spiral arms are nearly identical to those obtained without spiral arms, with average values of $6.19\,{\rm kpc}$ and $6.14\,{\rm kpc}$, respectively.
For individual stars, the differences in the mean radii between the barred models with and without spiral arms are only at the level of $\sim 0.01\,{\rm kpc}$.
The time evolution in Fig.~\ref{fig:rgc_time_potential_comparison} likewise shows that the spiral arms do not qualitatively alter the radial oscillation pattern over the integration time considered here.

Overall, this comparison shows that the non-axisymmetric bar potential has a significant impact on the inferred radial extent of the stellar orbits.
Compared to the axisymmetric case, the barred potentials produce smaller mean and median Galactocentric radii and allow the stars to reach substantially smaller pericentric radii.
However, the $R_{\rm GC}(t)$ curves remain strongly oscillatory rather than showing a monotonic radial drift.
}

\begin{table}
  \renewcommand\thetable{B.2}
\centering
\caption{ 
Mean, median, minimum, and maximum Galactocentric radii for the six stars in the axisymmetric, barred, and barred plus spiral-arm potentials.
}
\label{tab:rgc_statistics}
\begin{tabular}{llrrrr}
\toprule
 &  & 
$\langle R_{\rm GC} \rangle$ &
$\widetilde{R}_{\rm GC}$ &
$R_{\rm GC,min}$ &
$R_{\rm GC,max}$ \\
\cline{3-6}
\rb{Potential} & \rb{Hip} & 
\multicolumn{4}{c}{[kpc]} \\
\midrule
\midrule
McMillan17 & 11575 & 6.938 & 7.054 & 5.441 & 8.193 \\
McMillan17 & \phantom{1}9353  & 7.096 & 7.186 & 5.772 & 8.236 \\
McMillan17 & 22336 & 6.880 & 7.080 & 4.829 & 8.502 \\
McMillan17 & 22940 & 7.066 & 7.174 & 5.646 & 8.266 \\
McMillan17 & 23311 & 7.098 & 7.192 & 5.750 & 8.253 \\
McMillan17 & 26834 & 6.941 & 7.077 & 5.317 & 8.279 \\
\midrule
H24 noSpiral & 11575 & 6.196 & 6.102 & 3.694 & 8.480 \\
H24 noSpiral & \phantom{1}9353  & 6.209 & 6.130 & 3.665 & 8.612 \\
H24 noSpiral & 22336 & 6.184 & 6.244 & 3.316 & 8.614 \\
H24 noSpiral & 22940 & 6.218 & 6.124 & 3.625 & 8.644 \\
H24 noSpiral & 23311 & 6.218 & 6.134 & 3.647 & 8.641 \\
H24 noSpiral & 26834 & 6.196 & 6.115 & 3.626 & 8.528 \\
\midrule
H24 plusSpiral & 11575 & 6.184 & 6.089 & 3.700 & 8.431 \\
H24 plusSpiral & \phantom{1}9353  & 6.195 & 6.136 & 3.723 & 8.513 \\
H24 plusSpiral & 22336 & 6.182 & 6.232 & 3.345 & 8.593 \\
H24 plusSpiral & 22940 & 6.208 & 6.123 & 3.652 & 8.574 \\
H24 plusSpiral & 23311 & 6.206 & 6.144 & 3.682 & 8.557 \\
H24  plusSpiral & 26834 & 6.183 & 6.108 & 3.657 & 8.453 \\
\bottomrule
\end{tabular}
\end{table}

\end{document}